\documentclass{aa}
\usepackage{graphicx}
\usepackage{txfonts}
\usepackage{longtable,lscape}
\usepackage{natbib}

\newcommand{\cd}{d$^{-1}$\xspace}

\defcitealias{huemmerich20}{Paper 1}

\begin{document}

   \title{Photometric variability of the LAMOST sample of magnetic chemically peculiar stars as seen by TESS\thanks{Table A.1 is available in electronic form at the CDS via anonymous ftp to cdsarc.cds.unistra.fr (130.79.128.5) or via https://cdsarc.cds.unistra.fr/cgi-bin/qcat?J/A+A/}}

   \titlerunning{TESS view of mCP stars}

\author{J.~Labadie-Bartz\inst{1,2,3}
          \and
          S.~H{\"u}mmerich\inst{4,5}
          \and
          K.~Bernhard\inst{4,5}
          \and
          E.~Paunzen\inst{6}
          \and
          M.~E.~Shultz\inst{7}
          }

\institute{LESIA, Paris Observatory, PSL University, CNRS, Sorbonne University, Universit\'e Paris Cit\'e, 5 place Jules Janssen, 92195 Meudon, France
          \and
Instituto de Astronomia, Geof{\'i}sica e Ci{\^e}ncias Atmosf{\'e}ricas, Universidade de S{\~a}o Paulo, Rua do Mat{\~a}o 1226, Cidade Universit{\'a}ria, B-05508-900 S{\~a}o Paulo, SP, Brazil
        \and
Homer L.\ Dodge Department of Physics and Astronomy, University of Oklahoma, 440 W. Brooks Street, Norman, OK 73019, USA
          \and
Bundesdeutsche Arbeitsgemeinschaft f{\"u}r Ver{\"a}nderliche Sterne e.V. (BAV), Berlin, Germany 
          \and
American Association of Variable Star Observers (AAVSO), Cambridge, USA 
          \and
Faculty of Science, Masaryk University, Department of Theoretical Physics and Astrophysics, Kotl\'{a}\v{r}sk\'{a} 2, 611\,37 Brno,  Czechia
        \and
Department of Physics and Astronomy, University of Delaware, 217 Sharp Lab, Newark, Delaware 19716, USA}

   \date{}

  \abstract
    {High-quality light curves from space-based missions have opened up a new window on the rotational and pulsational properties of magnetic chemically peculiar (mCP) stars and have fuelled asteroseismic studies. They allow the internal effects of surface magnetic fields to be probed and numerous astrophysical parameters to be derived with great precision.}
   {We present an investigation of the photometric variability of a sample of 1002 mCP stars discovered in the Large Sky Area Multi-Object Fiber Spectroscopic Telescope (LAMOST) archival spectra with the aims of measuring their rotational periods and identifying interesting objects for follow-up studies.}
   {Transiting Exoplanet Survey Satellite (TESS) data were available for 782 mCP stars and were analysed using a Fourier two-term frequency fit to determine the stars' rotational periods. The rotational signal was then subtracted from the light curve to identify additional non-rotational variability signals. A careful pixel-level blending analysis was performed to check whether the variability originates in the target star or a nearby blended neighbour. We investigated correlations between the observed rotational periods, fractional age on the main sequence, mass, and several other observables.}
   {We present rotational periods and period estimates for 720 mCP stars. In addition, we have identified four eclipsing binary systems that likely host an mCP star, as well as 25 stars with additional signals consistent with pulsation (12 stars with frequencies above 10 \cd\ and 13 stars with frequencies below 10 \cd). We find that more evolved stars have longer rotation periods, which is in agreement with the assumption of the conservation of angular momentum during the main-sequence evolution.}
   {With our work, we increase the sample size of mCP stars with known rotation periods and identify prime candidates for detailed follow-up studies. This enables two paths towards future investigations: population studies of even larger samples of mCP stars and the detailed characterisation of high-value targets.}

   \keywords{stars: chemically peculiar --
                stars: rotation --
                techniques: photometric --
                stars: binaries: eclipsing --
                stars: oscillations (including pulsations)
               }

   \maketitle

\section{Introduction}

The chemically peculiar (CP) stars of the upper main sequence form a significant fraction of the upper main-sequence stars (about 10 per cent) and are encountered between spectral types early B to early F. Their defining characteristic is the presence of spectral peculiarities that indicate unusual elemental abundance patterns \citep[e.g.][]{preston74,maitzen84,smith96,gray09,ghazaryan18}, which are thought to originate from the interplay between radiative levitation and gravitational settling taking place in the calm outer layers of slowly rotating stars \citep[atomic diffusion; e.g.][]{michaud70,richer00}.

Following \citet{preston74}, the four main groups of CP stars are the CP1 stars (the metallic-line or Am/Fm stars), the CP2 stars (the magnetic Bp/Ap stars), the CP3 stars (the mercury-manganese or HgMn stars), and the CP4 stars (the He-weak stars). Although the observed abundance patterns within a group can vary considerably, each group is characterised by a distinct set of peculiarities. The CP1 stars exhibit under-abundances of Ca and Sc and over-abundances of iron-peak and heavier elements. The main characteristics of the CP2 stars are excesses of elements such as Si, Sr, Eu, or the rare-earth elements. Some of the most peculiar objects belong to this group, such as the extreme lanthanide star HD\,51418 \citep{jones74} or Przybylski's Star, HD\,101065 \citep{przybylski66}. The CP3 stars show enhanced lines of Hg and Mn and other heavy elements, whereas the CP4 stars possess anomalously weak He lines. Additional classes of CP stars have been proposed, for example the $\lambda$ Bootis stars \citep{gray97a,paunzen04,murphy17}, which are characterised by unusually low surface abundances of iron-peak elements, or the He strong stars, which are early B stars that exhibit anomalously strong He lines in their spectra \citep{bidelman65,morgan78}. As regards the strength of chemical peculiarities, a continuous transition from chemically normal to CP stars is observed \citep{loden87}.

Several authors have divided the CP stars into a `magnetic' and a `non-magnetic' sequence \citep[e.g.][]{preston74,maitzen84}. The former is made up of the CP2 and the He-peculiar stars (i.e. the CP4 and the He-strong stars), which possess strong and stable magnetic fields, while the latter encompasses, for example, the CP1, CP3, and $\lambda$ Bootis stars. While this canonical view has been frequently challenged (cf. e.g. \citealt{hubrig10,hubrig12,kochukhov13}, and \citealt{hubrig20} on the ongoing controversy about the presence of weak and tangled magnetic fields in CP3 stars), the groups of stars of the non-magnetic sequence certainly lack the strong and organised magnetic fields observed in the CP2 and He-peculiar stars, which can attain strengths of up to several tens of kilogauss \citep{babcock47,auriere07}. For convenience, CP2 and He-peculiar stars are generally referred to as magnetic chemically peculiar (mCP) stars -- a convention that we adhere to in this study.

While the origin of their magnetic fields is still a matter of some controversy \citep{moss04}, evidence has been collected in favour of the fossil field theory \citep[e.g.][]{braithwaite04}, according to which the magnetic field is a relic of the interstellar magnetic field that was `frozen' into the stellar plasma during star formation. Alternatively, a fossil field may be generated during a merger event \citep[e.g.][]{tutukov10,Schneider2019}.

Magnetic CP stars show a non-uniform surface distribution of chemical elements, which is associated with the presence of the magnetic field and manifests itself in the formation of spots and patches of enhanced or depleted element abundance. Flux is redistributed in these `chemical spots' (line and continuum blanketing; e.g. \citealt{wolff71,molnar73,lanz96,shulyak10,krticka13}), and mCP stars show strictly periodic light, spectral, and magnetic variations with the rotation period, which are satisfactorily explained by the oblique-rotator model (the magnetic axis is oblique to the rotation axis; \citealt{stibbs50}). According to convention, photometrically variable mCP stars are referred to, after their bright prototype, as $\alpha^{2}$ Canum Venaticorum (ACV) variables \citep{GCVS}. Rotation periods generally range from about 0.5 days to decades, with a peak at $\sim$2 days \citep[e.g.][]{RM09,bernhard15a}.

In addition to the rotational light changes, mCP stars can also exhibit pulsational variability. For a long time, the only proven form of pulsational variability among these objects was observed in the so-called rapidly oscillating Ap stars \citep{kurtz82}, which exhibit variability in the period range 5$-$20 min (high-overtone, low-degree, and non-radial pulsation modes). With the advent of ultra-precise space photometry, additional pressure modes (p modes) and gravity modes (g modes) associated with $\gamma$ Doradus \citep[e.g.][] {kaye99,guzik00} and $\delta$ Scuti \citep[e.g.][]{breger00} pulsations were identified in a number of CP2 stars \citep[e.g.][]{balona11,cunha19,holdsworth21}. In general, the high-quality light curves (LCs) from space-based missions open up the possibility for the asteroseismic characterisation of mCP stars, which in turn allows the internal effects of surface magnetic fields to be probed and numerous astrophysical parameters to be derived with great precision \citep{briquet12,buysschaert18}.

The most up-to-date collection of CP stars is the General Catalogue of CP Stars \citep{RM09}, which was published more than a decade ago. It lists about 3500 mCP stars or candidates ($\sim$2000 confirmed mCP stars and $\sim$1500 candidates) and is still one of the main resources regularly employed in investigations of mCP stars. However, in recent years, several studies have published new samples of these objects (e.g. \citealt{huemmerich18,scholz19,sikora19}). Of note is the study of \citet[][hereafter Paper 1]{huemmerich20}, who published a sample of 1002 mCP stars, thereby significantly enlarging the total known number of these objects.

Here we present our efforts to characterise the photometric variability of the sample of mCP stars published in \citetalias{huemmerich20} using photometric time-series observations from the NASA Transiting Exoplanet Survey Satellite \citep[TESS;][]{Ricker2015}. Our paper is structured as follows. The employed data sources and methods are described in Sect.~\ref{sec:methods_and_data}. In Sect.~\ref{sec:discussion} we present and discuss our results. Special emphasis is placed on the eclipsing binary (EB) systems and pulsator candidates in our sample, which are studied in detail in Sect.~\ref{sec:special_objects}. We conclude in Sect.~\ref{sec:conclusion}.

\section{Methods and data}
\label{sec:methods_and_data}

\subsection{The LAMOST DR4 sample of mCP stars}
\label{sec:sample_of_mCP_stars}

In our search for mCP stars \citepalias{huemmerich20}, we employed a modified version of the \textsc{MKCLASS} code\footnote{\url{http://www.appstate.edu/~grayro/mkclass/}}, a computer program conceived by Richard O. Gray to classify stellar spectra on the MK system, to search for mCP stars in spectra from the fourth data release (DR4) of the Large Sky Area Multi-Object Fiber Spectroscopic Telescope (LAMOST) of the Chinese Academy of Science \citep{lamost1,lamost2}. LAMOST is a Schmidt telescope based at Xinglong Observatory (Beijing, China) that boasts an effective aperture of 3.6$-$4.9\,m (field of view of about 5$\degr$) and is able to collect 4000 spectra in a single exposure (spectral resolution R\,$\sim$\,1800, limiting magnitude $r$\,$\sim$\,19\,mag, wavelength coverage 3700 to 9000\,\AA). LAMOST is therefore perfectly suited for large-scale spectral surveys; data products are made available to the public in consecutive data releases accessible through the LAMOST spectral archive.\footnote{\url{http://www.lamost.org}}

In a nutshell, suitable candidates were collected from a colour-selected sample of early-type stars by searching for the presence of the 5200\,\AA\ flux depression, a characteristic of mCP stars \citep[e.g.][]{kodaira69,maitzen76,paunzen05,kochukhov05,khan06}. Spectral classification was then performed using MKCLASS\_mCP, a version of the original program modified to probe a number of spectral features relevant to the identification and classification of mCP stars. In this way, a final sample of 1002 mCP stars (mostly CP2 stars and several CP4 stars) was collected, most of which were new discoveries (only 59 objects have an entry in the \citet{RM09} catalogue). These objects are between 100\,Myr and 1\,Gyr old, with the majority having masses between 2\,$M_\odot$ and 3\,$M_\odot$. From an investigation of a sub-sample of 903 mCP stars with accurate astrophysical parameters, we determine a mean fractional age on the main sequence of $\tau$\,=\,63\,\% (standard deviation of 23\,\%) and conclude that our results provide evidence for an inhomogeneous age distribution among low-mass ($M$\,$<$\,3\,M$_\odot$) mCP stars. For more detailed information on the methods employed, we refer the reader to \citetalias{huemmerich20}.

\subsection{TESS data} \label{sec:TESS}

The NASA TESS mission was launched in 2018 with the primary goal of discovering transiting exoplanets via high-precision time-series photometry. The four identical cameras of TESS cover a combined field of view of 24$^{\circ}$ $\times$ 96$^{\circ}$ and are pointed at a given region of the sky for 27.4 days (one observing sector). In its first two years of operation (the primary mission), TESS observed nearly the entire southern ecliptic hemisphere over 13 observing sectors (1 year), followed by a similar strategy for the northern ecliptic hemisphere. Thus, TESS has observed nearly the entire sky and continues to do so (while filling in small gaps in sky coverage) in its ongoing extended mission. TESS records red optical light with a wide bandpass spanning roughly 600$-$1000 nm, centred on the traditional Cousins $I$ band. For optimal targets, the noise floor is approximately 60 ppm h$^{-1}$.

The full frame images (FFIs) from TESS are made publicly available for its entire field of view. Therefore, LCs can be extracted for virtually any object observed by the satellite. During the primary mission (Cycles 1 and 2), the FFIs were delivered with a 30-minute cadence, then a 10-minute cadence for Cycles 3 and 4, and a 200-second cadence for Cycle 5 (the current cycle at the time of this writing). It is these FFIs that constitute the fundamental data products used in this work. TESS also provides 2-minute cadence LCs for pre-selected targets, but the majority of our mCP sample was not observed in this mode. At the time of writing, TESS data were available
for 782 of the 1002 mCP stars from \citetalias{huemmerich20}.

\subsection{Light curve extraction from TESS FFIs} \label{sec:LC_extract}

Light curves were extracted from the TESS FFIs for all of the stars in the sample observed by TESS, up to and including TESS sector 35 (the latest available sector at the time of the LC extraction). The \textsc{lightkurve} \citep{Lightkurve2018} and \textsc{TESScut} \citep{Brasseur2019} packages were used to download a 40 $\times$ 40 pixel target pixel file (TPF) centred on the coordinates of the target star. To determine the aperture used to generate the LC (via simple aperture photometry), an initial threshold of 10 sigma relative to the median flux level in the TPF selected an initial pixel mask centred on the target. The aperture size was then further constrained depending on the target brightness, being restricted to a radius of 2 pixels for the faintest targets ($T_{\rm mag}$ $\geq$ 11), and up to 5.5 pixels for the brightest ($T_{\rm mag}$ $\sim$5). Two different detrending methods were employed to remove systematic trends. The first involved simple background subtraction, where background pixels were identified in the TPF, and their average flux level in each frame was subtracted from the target LC (after accounting for the number of pixels in the adopted aperture). The second method first excluded pixels in a 10 $\times$ 10 square centred on the target (or larger for the brightest stars), and then the remaining pixels in the TPF were used as regressors in a principal component analysis (PCA) correction, using five PCA components to remove common trends across this region of the CCD. Both detrending methods have advantages and disadvantages; for instance, the PCA detrending can over-fit the data and remove genuine longer-term trends, while background subtraction can perform poorly for certain types of systematics (e.g. associated with spacecraft momentum dumps). Therefore, for each target, the `best' version of the LC (PCA versus background-corrected) was selected as described in Sect.\,\ref{sec:preprocessing}. In practice, the PCA detrending was preferred about 85\% of the time. For a small number of stars with problematic data from the initial LC extraction (e.g. when the star fell on the edge of the TESS CCD), LCs were later re-extracted to make use of the most recent TESS data available.

\subsection{Pre-processing the light curves} \label{sec:preprocessing}

The primary two pre-processing steps were to determine the single best version of the LC to use for a given star (PCA versus background-corrected) and to remove outliers. These were both done in a single routine as follows. Both LC versions were subjected to an initial multi-term Fourier analysis that aimed to fit frequencies below $\sim$0.5 \cd, which include rotational and slower systematic signals. This multi-term fit was subtracted from the LC, and from these residuals statistical outliers were identified automatically (being five standard deviations from the mean). At this stage, any additional outliers were manually selected via an interactive Python routine (or in some cases, statistical outliers were selected to remain if, e.g., the initial multi-term Fourier fit poorly reproduced the slower variability). The scatter in the (outlier-removed) residuals was compared for both versions of the LC, and typically the one with the lowest scatter was selected as the best version to use in the subsequent analysis. The original LC, with the low-frequency signals intact but without outliers, was then saved for further analysis. A simple sigma clipping to the original LCs was not optimal for outlier removal, since the astrophysical rotational signals were often of a higher amplitude than the deviations of the outliers from the mean flux.

\subsection{Variability analysis} \label{sec:var_analysis}

\begin{figure}
   \centering
   \includegraphics[width=1.0\hsize]{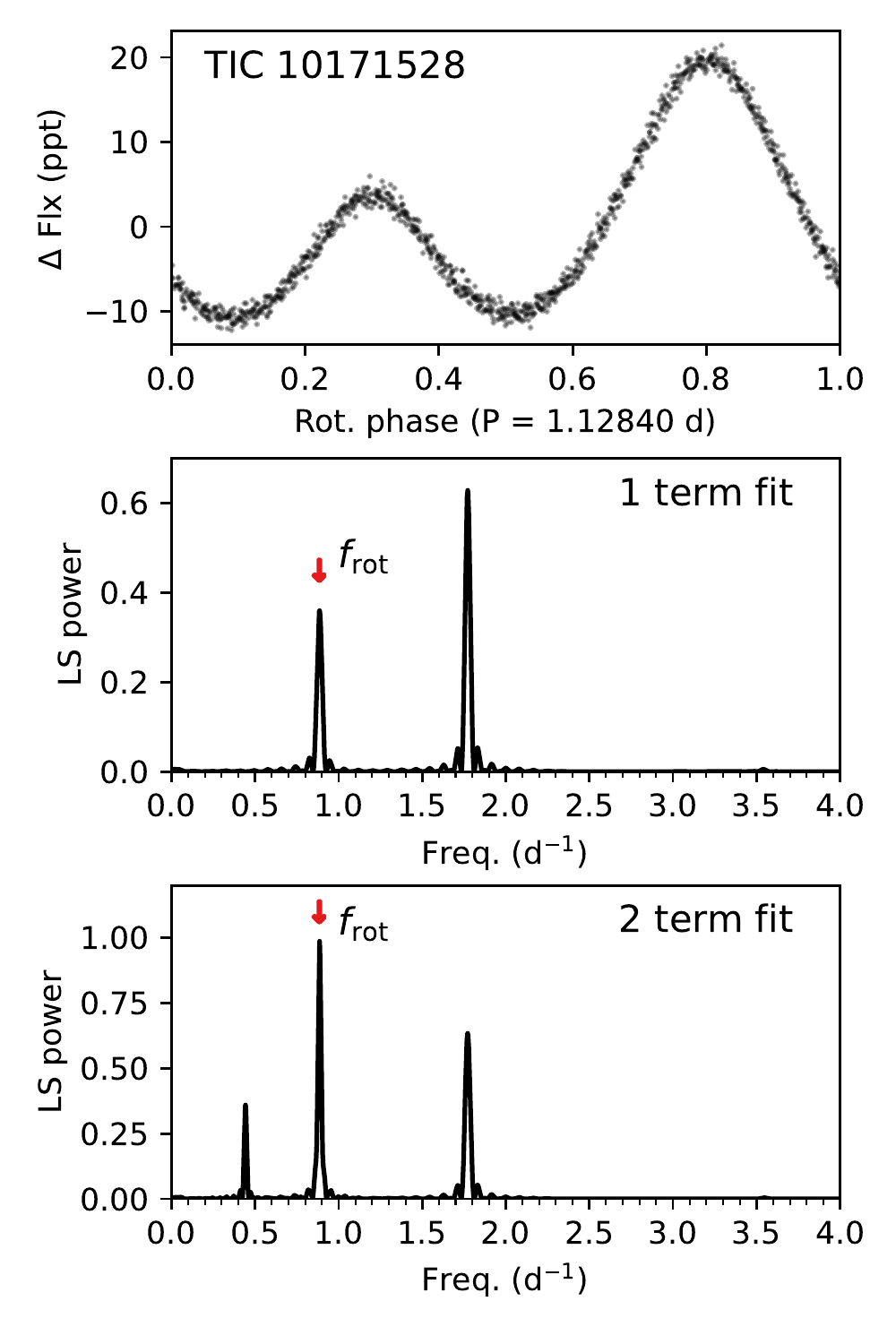}
      \caption{Comparison of one- and two-term Fourier analysis for identifying the rotation period. \textit{Top:} Phased TESS LC for an example mCP star with a decidedly double-waved pattern at the rotational period (i.e. two maxima and two minima per rotation). \textit{Middle:} Standard single-term frequency analysis, showing that the strongest peak is at 2$\times f_{\rm rot}$. \textit{Bottom:} Two-term frequency analysis, where the strongest peak is at $f_{\rm rot}$.
              }
         \label{fig:1vs2term_fit}
   \end{figure}

As expected for mCP stars, a preliminary analysis of the TESS LCs showed that rotational variability was by far the most common and dominant photometric signal for the sample. For stars with stable bright surface spots (e.g. mCP stars), the observed brightness in broadband photometry is modulated at the rotational period. The exact shape of the variability pattern depends on factors such as the inclination angle and the spot sizes and distribution, but in general the photometric signal is non-sinusoidal and thus forms a harmonic series in the frequency spectrum computed from the LC. With a standard Fourier analysis, frequency peaks are often found at the rotational frequency and its harmonics, but the strongest peak may be at one of these harmonics. Instead, a two-term frequency analysis (fitting $f$ and 2$\times f$ for a pre-defined grid of frequencies) more reliably found the strongest peak at the rotational frequency automatically and was thus preferred for the automated determination of rotational frequencies. An example of this is illustrated in Fig.\,\ref{fig:1vs2term_fit}. Including additional terms did not improve results. The primary tool we used to determine rotation periods was thus a modified generalised Lomb-Scargle periodogram \citep{Zechmeister2009,Press1992,VanderPlas2012,VanderPlas2015}; we used two Fourier terms, employing the \textsc{timeseries.LombScargle} package of \textsc{Astropy} \citep{astropy2013, astropy2018}.

This two-term frequency analysis was applied to the entire sample, and the strongest peak was presumed to be the rotational frequency and was thus tabulated. For each star, plots were made phasing the LC to 0.5$\times$, 1$\times$, and 2$\times$ this frequency and were manually inspected (along with the one- and two-term frequency spectrum) to ensure the correct rotational period was identified. All systems where the automatic analysis was in doubt were analysed manually with Period04 \citep{Lenz2005}. The most common reason for the automatic analysis to fail were cases where relatively strong systematic effects dominated the LC (especially for the faintest sources), but where rotational modulation could usually still be recovered by detrending against these (often much slower) systematics. For the slowest rotators, the short 27 d baseline of TESS was insufficient to sample a full rotational cycle and rotational periods could only be coarsely estimated (cf. also Sect.~\ref{sec:results}). In some cases, it was not possible to determine or even estimate a rotational period from the TESS data (e.g. when the rotation period was too long, the amplitude too small, or the data problematic).

A given star in our sample may also display non-rotational photometric variability due to, for example, stochastic low-frequency excess \citep[astrophysical correlated red noise,][]{Bowman2019,Bowman2020}, pulsation, binarity, or a combination thereof. To investigate this, a two-term model of the rotational modulation was subtracted from the LC, whereafter the standard (one-term) frequency spectrum was calculated, which was then fit to determine the red noise profile \citep[similar to the method used by][]{Bowman2020,Bartz2021}. The red noise profile, multiplied by 5, was then used as a threshold above which other statistically significant frequencies (in general not associated with the stellar rotation from the first step) were automatically identified as candidates with additional variability. This step led to the identification of 107 such candidates. 

However, given the large pixel size of TESS (21 arcseconds), contaminating flux from neighbouring stars falling into the aperture (`blending') is a concern. Therefore, LCs were re-extracted for these candidates including up to the most recent Cycle 5 images for a pixel-level blending analysis. This was done in two steps. The first `coarse' analysis involved marking the location of all nearby \textit{Gaia} sources (with $G_{\rm mag} < 15$) overlaid on an image of the TESS TPF, and plotting the LC and frequency spectrum of each pixel in the vicinity of the target and then comparing the per-pixel frequency spectrum to that calculated from the PCA-detrended LC for the target star. In this way, candidate pulsational signals (or other variabilities) could be localised to determine if they originate in or near the target star or in a nearby blended neighbour. A suite of plots were inspected manually to make this determination. This analysis was generally sufficient to identify variable sources $\sim$2 or more pixels away from the mCP target star. Each object that passed this first analysis was then examined in more detail using the {\sc TESS\_localize}\footnote{\url{https://github.com/Higgins00/TESS-Localize}} Python package \citep{Higgins2022}. {\sc TESS\_localize} is designed to locate the origin of variability signals to within one-fifth of a TESS pixel. In all objects analysed with {\sc TESS\_localize}, the rotational variability was confirmed to originate in the mCP star, and two objects with additional frequencies were rejected as blends (in both these cases the blended variable was faint, $G_{\rm mag} \sim$17 to 18, and about one pixel away from the mCP star).

Each star determined to potentially have inherent variability in addition to rotation is listed in Sect.\,\ref{sec:special_objects}. Those objects with additional variability due to contaminating flux from blended neighbouring sources are indicated (see Sect.\,\ref{sec:results}) in order to minimise duplication of efforts in future studies. We note that the typically low-amplitude signals from blended neighbours do not impact the analysis of the rotational variability of these objects -- no cases were identified where the presumed rotational modulation originates off-target.

\section{Results and discussion}
\label{sec:discussion}

\subsection{Presentation of results}
\label{sec:results}

Table~\ref{table_master1} in the appendix lists essential data for our sample stars, including identifiers from the TESS Input Catalogue (TIC; \citealt{TIC}) and LAMOST; positional information and magnitude in the $G$\,band from \textit{Gaia} DR3 \citep{GAIA1,GAIA2,GAIA3}; spectral type from \citetalias{huemmerich20}; and variability period and peak-to-peak amplitude, as derived from TESS data in this study. Where appropriate, additional information is provided in the column `Remarks', which includes variability information gleaned from the International Variable Star Index of the AAVSO (VSX; \citealt{VSX}). The VSX is the most up-to-date collection of variable star data and constantly updated with new variability catalogues and results from the literature. It contains the results of most relevant papers dealing with the periods of ACV variables (including more recent studies,  such as \citealt{paunzen98}, \citealt{wraight12}, \citealt{bernhard15a}, \citealt{bernhard15b}, \citealt{huemmerich16}, and \citealt{bernhard21}). Not included at the time of this writing are the studies of \citet{Bowman2018}, \citet{sikora19}, \citet{daviduraz19}, \citet{bernhard20}, \citet{2020A&A...639A..31M}, and \citet{2022A&A...660A..70M}. Except for two stars from the list of \citet{2022A&A...660A..70M}\footnote{\citet{2022A&A...660A..70M} identified TIC 239801694 and TIC 368073692 as candidate very slowly rotating mCP stars; no variations were reported. Our analysis also found no variations in TIC 239801694, but for TIC 368073692 we found a period of 20.49 d with a consistent signal in both available sectors of data. }, there are no matches between the samples of these studies and our sample.

The VSX has data for only 64 out of the 782 objects presented here. With the present work, we therefore significantly add to the sample of mCP stars with accurately determined rotational periods. We note that for several objects, the VSX designations are not accurate (although the periods may still be correct), for example when variability was interpreted as being related to binarity and not rotation or when the object was included under a generic variability type such as ROT or MISC. These objects are here identified as ACV variables for the first time.

\begin{figure}
   \centering
   \includegraphics[width=1.0\hsize]{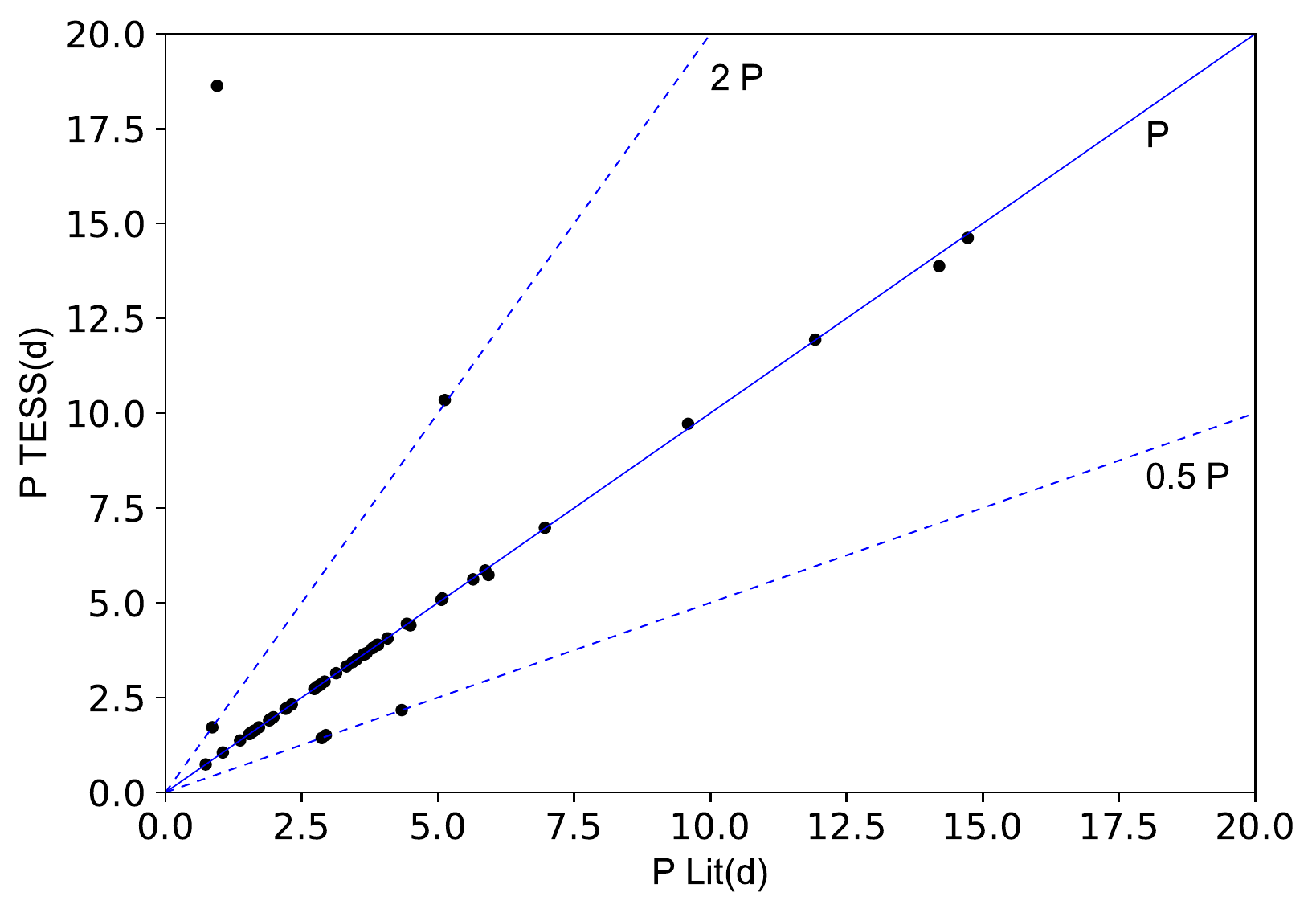}
      \caption{Comparison of literature periods and periods derived from the analysis of TESS data in this work for the 64 stars with periods listed in the VSX. The outlying data point refers to TIC 237662091, which is discussed in the text.}
         \label{fig:plitptess}
   \end{figure}

Figure \ref{fig:plitptess} illustrates a comparison of the literature periods from the VSX with the periods derived from TESS data in this work. In five cases, the period values in the VSX obviously represent half (TIC 235273014, TIC 2679119) or twice (TIC 438165498, TIC 250478934, TIC 21018674) the true rotation period. This is expected, as double-waved ACV LCs can easily be misinterpreted as single-waved LCs (or vice versa) if the data are noisy, the amplitude of the light variations is very small, or the difference in brightness between both maxima or minima is negligible. In this respect, the ultra-precise TESS data clearly have an advantage over ground-based photometric data sources for typical periods. Apart from that, the agreement is very good, which highlights the quality of the VSX period data. The single exception is for TIC 237662091, where TESS clearly revealed a long rotation period (18.630 d, although it is possible the true rotation period is twice this value), and the VSX period corresponds to the daily alias (0.94661 d).
 
   \begin{figure*}
   \centering
   \includegraphics[width=\hsize]{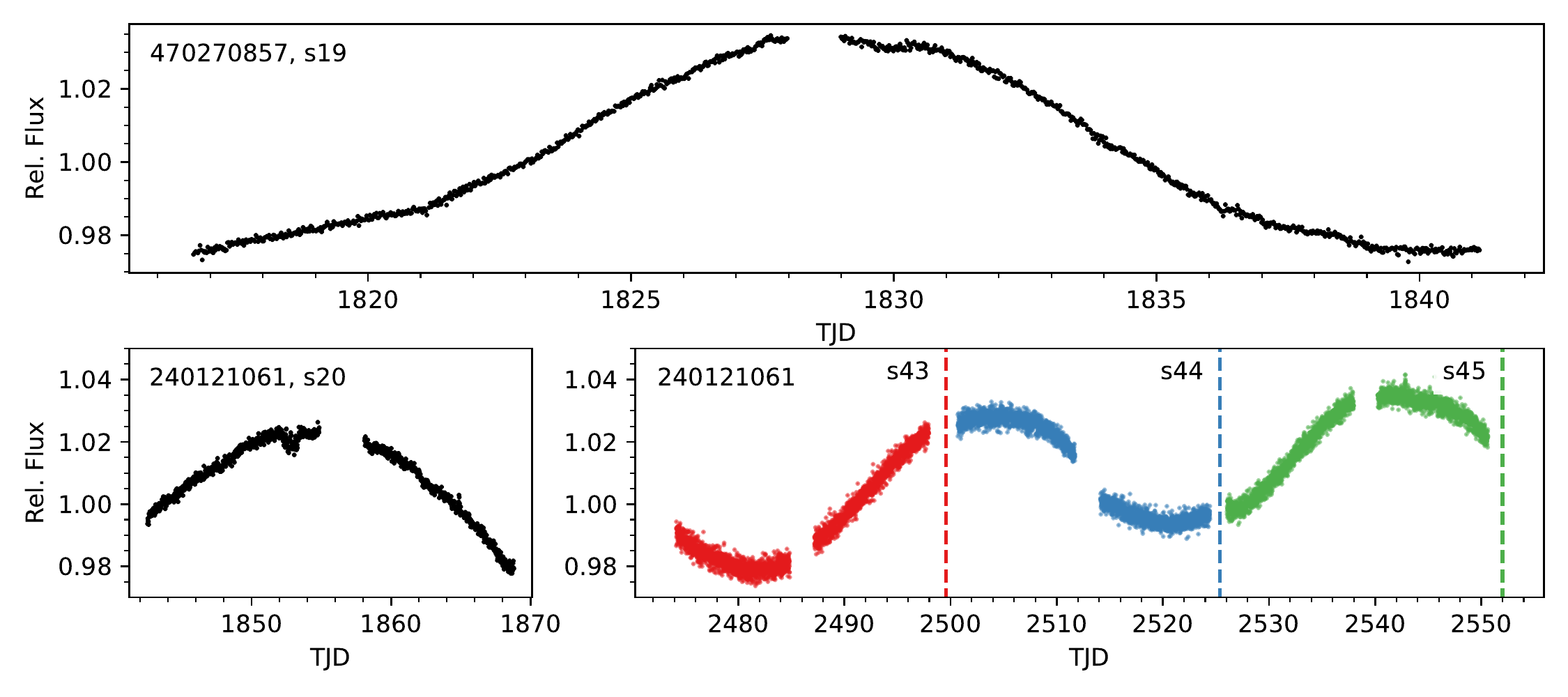}
      \caption{Two stars with periods longer than a single TESS sector. For TIC 470270857 (top), only one sector of data was available. For TIC 240121061, one sector in Cycle 2 and three consecutive sectors in Cycle 4 were available. Even with three sectors, the rotation period is difficult to determine -- it may be $\sim$40 days (if single-waved) or $\sim$80 days (if double-waved).}
         \label{fig:longP}
   \end{figure*}

Periods are given to the last significant digit. Approximate periods are identified by the use of a colon (`:'). Several stars are clearly variable but have periods too long to be resolved with the rather short time baseline of the employed TESS data. That is expected, as ACV variables can have periods of up to decades or even longer \citep{mathys19, 2019A&A...624A..32M, 2020A&A...636A...6M, 2020A&A...639A..31M, 2022A&A...660A..70M}. These objects are identified by the remark `long-period var.' in Table~\ref{table_master1}. We caution that it is not always straightforward to distinguish between long-term trends inherent to the TESS data and intrinsic stellar variability, in particular when the data are noisy. Therefore, a question mark is added to the aforementioned remark when the situation remains unclear.

Most of these suspected long-period ACV variables boast data from one sector only (time baseline of 27 days) so that not more than a part of the rotational cycle was covered in the available data. These objects are consequently listed with a period of ``>30:'' in Table~\ref{table_master1}. Two examples of stars with periods longer than a single TESS sector are shown in Fig.~\ref{fig:longP}, which also illustrates the difficulties encountered in period determination for these objects even when data from several sectors are available. Since each sector is reduced separately, there are different systematic trends and different amounts of contaminating flux from neighbouring stars, so the measured photometric amplitude can differ from one sector to the next, and different constant flux offsets may need to be applied to each sector of data (or even each half-sector; to avoid discontinuities). For stars such as these, the background-corrected version of the LC typically produces better results (as plotted here) since the PCA detrending tends to remove or distort the longer-term signals (see Sect.\,\ref{sec:LC_extract}).

In 62 stars, no variability could be inferred from TESS data. For 23 of these objects, a LC of reasonable or good quality is extracted, yet no variability is seen (remark `data fine, no var.'). We consider these stars prime candidates for very slowly rotating mCP stars. One of these objects (TIC 239801694) has been proposed to be a very slowly rotating mCP star candidate by \citet{2022A&A...660A..70M}.

For the remaining objects, blending issues or problems with either the LC extraction or detrending yield unreliable results. Some of these stars do not fall on useful pixels on the TESS CCD, and thus no LC could be extracted (remark `edge of CCD, no LC extracted'). Other targets are faint and in a relatively crowded field and so the automatic aperture selection failed (remark `faint, aperture selection failed'), or there is significant blending from (often relatively bright) neighbouring sources, rendering it difficult or impossible to isolate the target star (remark `blending dominates, LC unreliable'). Finally, in some LCs there are problems in detrending against systematic effects making it difficult to determine whether or not there is any astrophysical variability (`systematics dominate, LC unreliable'). 

Stars of special interest, which are further discussed in Sect.~\ref{sec:special_objects}, are marked by an asterisk (`*') in the column `Remarks'. These are the EBs (Sect.~\ref{subsec:eclisping_binaries}) and stars with additional signals that we attribute to pulsation (or in some cases where these additional signals have a harmonic structure to perhaps rotation or binarity of an unresolved object; Sect.~\ref{subsec:pulsators}). In regard to the pulsator candidates, corresponding remarks identify whether additional signals in the low-frequency ($<$10 \cd; `add. low-freq. signal') or high-frequency ($>$10 \cd; `add. high-freq. signal') realms were detected.

The LCs of several objects clearly show eclipses or other additional variability that a pixel-level blending analysis revealed to originate in a neighbouring star in close proximity on the sky (cf. Sects.~\ref{sec:var_analysis} and \ref{subsec:eclisping_binaries}). These objects are identified by the remark `eclipses not on target' and `add. var. not on target', respectively, in Table~\ref{table_master1}, mainly to avoid confusion in further studies dealing with these stars.

\subsection{Rotation period versus fractional age on the main sequence and stellar mass}
\label{sec:phot_var_properties}

\begin{figure}
\centering
\includegraphics[width=1.0\hsize]{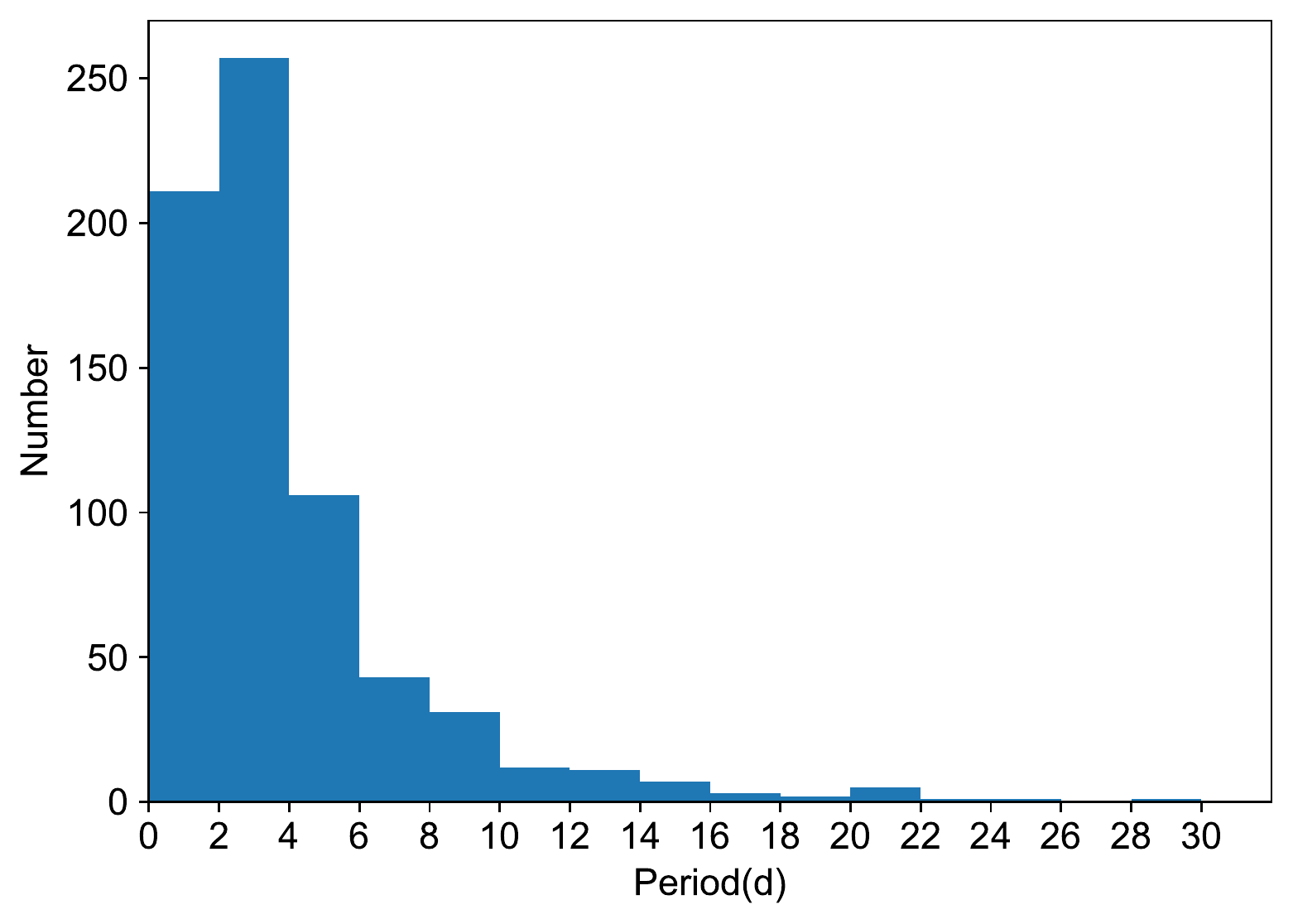}
\includegraphics[width=1.0\hsize]{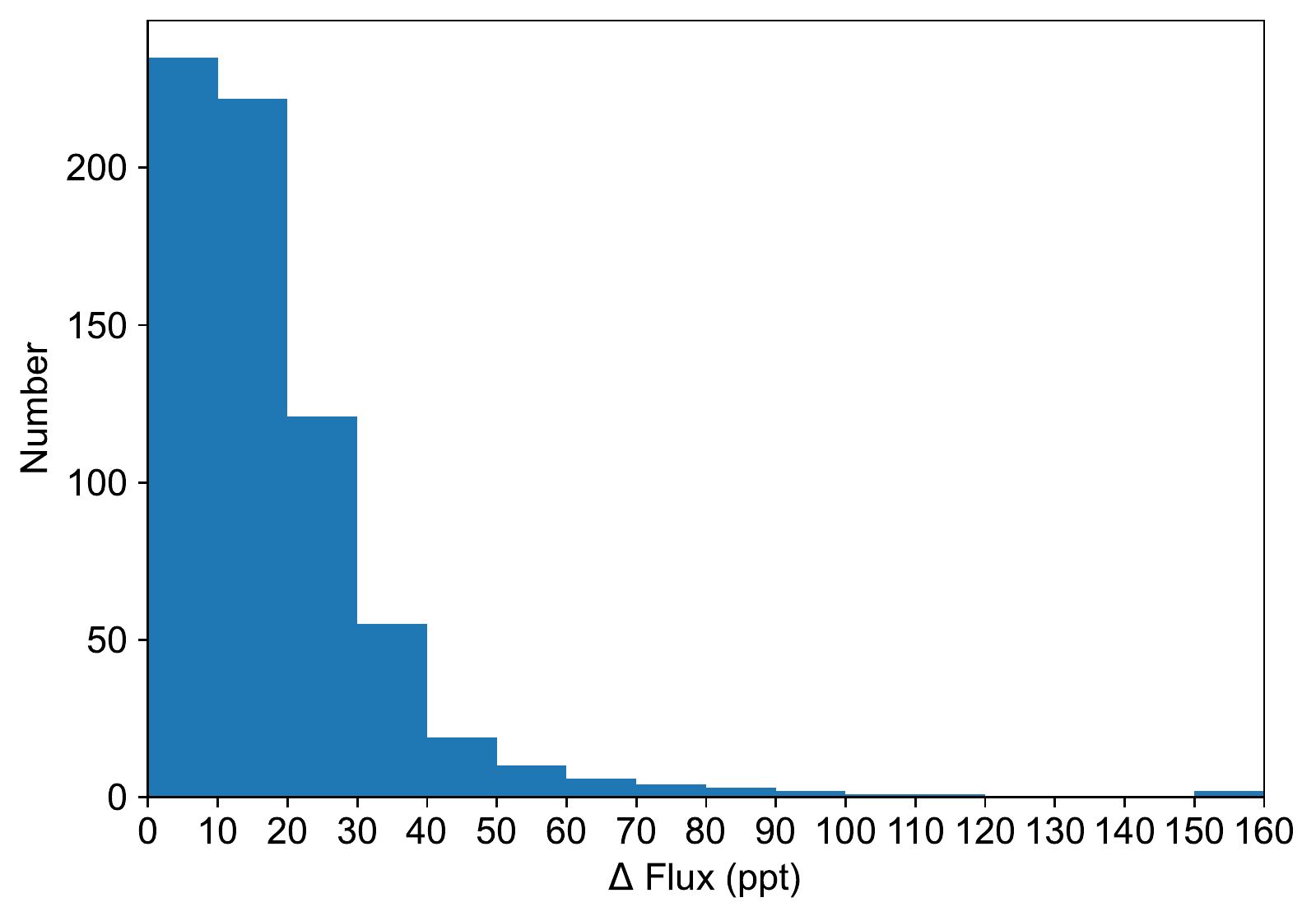}
   \caption{Histograms of the distribution of rotational periods (upper panel) and photometric peak-to-peak amplitudes in the TESS bandpass (lower panel) of the 720 sample stars for which these parameters could be derived.}
      \label{fig:histograms}
\end{figure}

   \begin{figure*}
   \centering
   \includegraphics[width=\hsize]{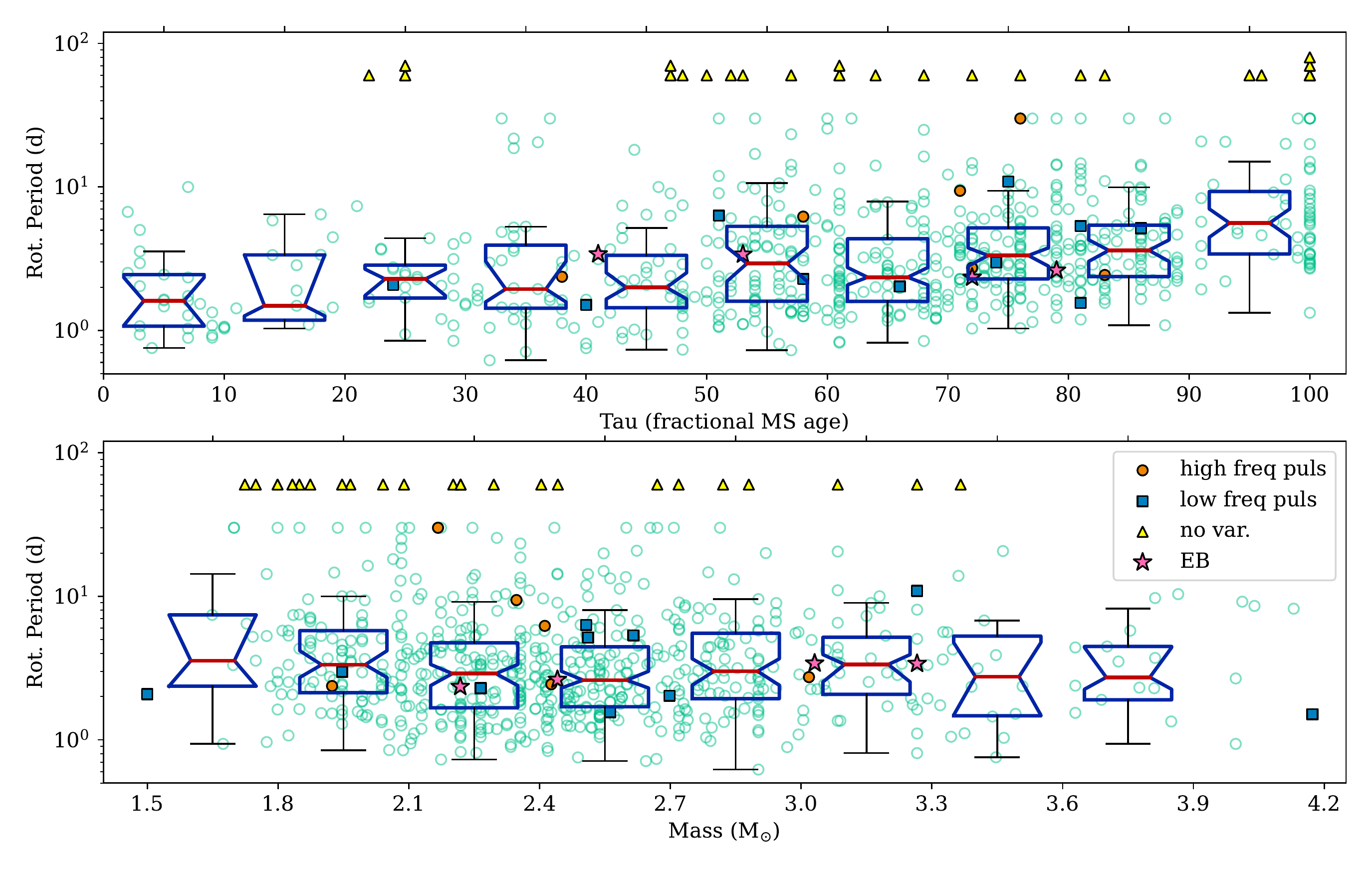}
      \caption{Derived rotation period from TESS versus the fractional main-sequence age (Tau, expressed as a percentage; upper panel) and stellar mass (lower panel). Data were binned, with the overlaid box plots extending from the lower to upper quartile values and the horizontal red line representing the median. The notches in the box indicate the 95\% confidence interval of the median, and the whiskers extend to the limits of the binned data, excluding the statistical outliers. In the bottom panel, the first bin includes masses less than 1.8 $M_{\odot}$, and the last bin all masses greater than 3.6 $M_{\odot}$. Otherwise, the mass bins are 0.3 $M_{\odot}$ wide. There is a correlation with more evolved stars having longer rotation periods, but there is apparently no correlation with the mass.  
      Stars with additional high-frequency ($>$10 \cd) or low-frequency ($<$10 \cd) signals are indicated by the orange circles and blue squares, respectively (not all of them have stellar parameters from \citetalias{huemmerich20}). Yellow triangles mark the 23 stars where no variability was detected (and we could not measure a rotation period), which are candidate very slow rotators. Stars with estimated rotation periods of $``>$30:'' (i.e. slow but detectable rotational variability longer than a single TESS sector, as in Fig.~\ref{fig:longP}) are included in this plot with $P_{\rm rot}$ = 30 d as a lower limit. The four EBs are marked by pink stars, but their mass and age may not be accurate since their binarity was not considered in \citetalias{huemmerich20}.}
         \label{fig:P_vs_TM}
   \end{figure*}

Figure \ref{fig:histograms} investigates the distribution of rotation periods (upper panel) and the photometric peak-to-peak amplitudes in the TESS bandpass (lower panel) of the 720 sample stars for which these parameters could be derived from TESS data. Our results are in excellent agreement with the literature and illustrate the well-known peak around $P$\textsubscript{rot} $\sim$ 2 days \citep[e.g.][]{RM09,bernhard15a,bernhard15b,huemmerich16,2019MNRAS.483.3127S} and the typical magnitude range of the photometric variations \citep[e.g.][]{manfroid86,bernhard15a} that appear somewhat reduced in the broad red bandpass of the TESS mission (600$-$1000 nm; cf. \citealt{bernhard20}).

The correlations between the derived rotation periods with the fractional age on the main sequence (Tau) and the stellar mass (M) are illustrated in Fig.~\ref{fig:P_vs_TM}. Tau and M values are taken from \citetalias{huemmerich20}. We find a correlation in the sense that more evolved stars have longer rotation periods, which is in agreement with results from the literature that confirmed that the evolution of rotation periods among mCP stars agrees very well with the assumption of conservation of angular momentum during the main-sequence evolution \citep[cf. e.g. the discussion in][]{Adelman2002,netopil17}. Magnetic braking is well known to spin down magnetic B stars as they evolve \citep{2019MNRAS.490..274S}, and could also be a factor in the trend observed in our mCP sample of slower rotation with age. However, it is generally suggested that magnetic braking is weak in mCP stars and that the primary reason for spin-down is changes of the moment of inertia \citep[with angular momentum conserved;][]{1998A&A...334..181N, 2006A&A...450..763K}.

Apparently, there is no correlation between rotation period and mass. The stars with no detected rotational variability (presumed to be very slow rotators) have a similar distribution of age compared with the majority of the sample where a rotation period is determined, but may preferentially have lower masses. However, with only 23 very slow rotator candidates, the sample is too small to draw any statistical conclusions.

While our results lend themselves perfectly for further statistical analyses, these are out of scope of this study and are best investigated with as large a sample of mCP stars as possible. This will be the topic of an upcoming paper (Paunzen et al., in preparation). Nevertheless, some preliminary findings on the correlations between several observables are shortly discussed in Sect.~\ref{sect:correlations}.

\section{Note on special objects}
\label{sec:special_objects}

\subsection{Eclipsing binaries}
\label{subsec:eclisping_binaries}

\begin{figure*}
   \centering
   \includegraphics[width=\hsize]{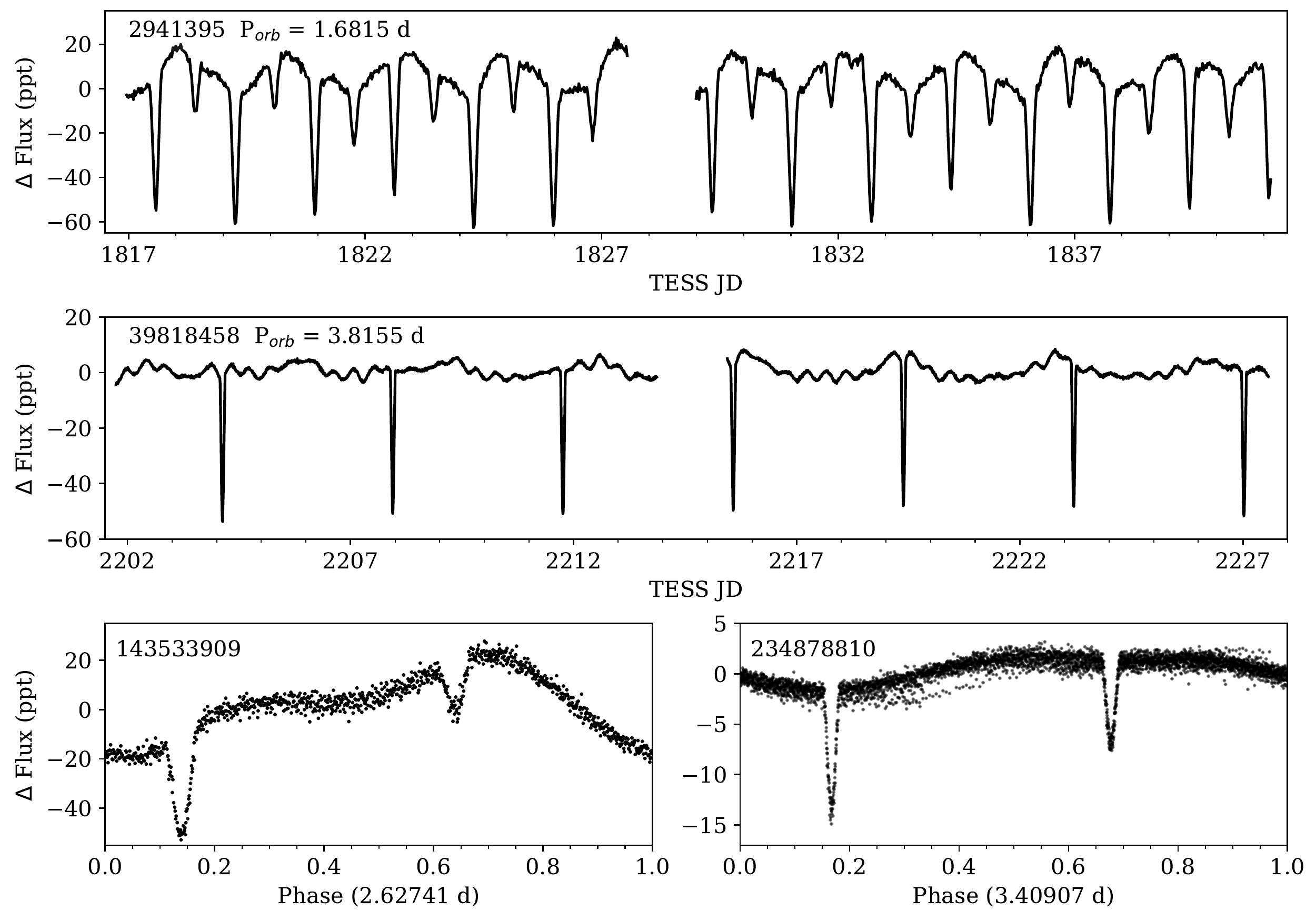}
      \caption{LCs of the four EBs identified in the sample. In the top two panels, a single sector of TESS data is shown, un-phased (since neither the rotational nor pulsational signals are synchronised to the binary orbit). The bottom two panels are phased to the rotational periods, which are identical to the orbital periods. TIC identifiers are indicated in each panel.
              }
         \label{fig:EBs}
   \end{figure*}

Four systems are found to be EBs. Many additional targets contain eclipses in the extracted LCs, but a pixel-level blending analysis (Sect.\,\ref{sec:var_analysis}) shows that the eclipses originate in a neighbouring star. These four EBs are listed and briefly described below and are shown in Fig.\,\ref{fig:EBs}. EBs containing CP stars are rare and are valuable opportunities for precise determinations of the stellar and system properties \cite[e.g.][]{Kochukhov2021}. 
Among the hot stars on the upper main sequence, global magnetic fields are found in about 10\% of effectively single stars, but only in 2\% of close binaries \citep{Alecian2015}. Discovering and characterising EBs which host magnetic stars are important for constraining the origin of magnetism in hot stars \citep[e.g.][]{Shultz2019}. 

TIC 2941395 (= UCAC4 624-023360, $V_{\rm mag}$ = 13.5): 
The rotational signal is not in phase with the eclipses -- that is, the rotation and binary periods are not identical, with the mCP star apparently rotating slower than the orbit. A frequency analysis was performed after clipping out all eclipses to derive a rotation period of $P_{\rm rot}$ = 2.346688 d ($f_{\rm rot}$ = 0.426132 c/d). The orbital period is $P_{orb}$ = 1.6815 d ($f_{orb}$ = 0.5947 c/d). At such short orbital periods, synchronisation is expected, perhaps casting doubt that the mCP star is involved in the eclipsing system as in, for example, a hierarchical triple system with an inner EB and an outer mCP star. This has not previously been reported as an EB.

TIC 39818458 (= HD 40759, $V_{\rm mag}$ = 9.3): 
This is an EB, with eclipses on-target, and is included in the TESS OBA EB catalogue of \citet{IJspeert2021}.
There appears to be variability caused by three different scenarios -- rotation, pulsation, and orbital. In addition to the clearly visible slower pulsation (consistent with g-modes), there are also many higher-frequency $\delta$ Scuti-like pulsation modes (between $\sim$50 -- 60 \cd). All eclipses are the same depth (i.e. there is no visible secondary eclipse) and are not flat-bottomed, occurring every 3.8155 days. The rotational period of the mCP star is determined to $P_{\rm rot}$ = 3.3949695 d. This system is being characterised in more detail by Semenko et al., in prep, and so is not discussed further in this work.

TIC 143533909 (= LAMOST J041819.79+414611.3, $G_{\rm mag} = 14.2$): 
This is an EB with a primary and secondary eclipse that are synchronised to the rotation period (2.6274 d). The eclipses occur near, but not exactly at, the maximum and minimum of the rotational brightness. With identical orbital and rotational periods, it seems likely that the mCP star is genuinely part of the EB system. This has not previously been reported as an EB.

TIC 234878810 (= HD 259273, $V_{\rm mag}$ = 9.73): 
This is an EB with primary and secondary eclipses (depths of about 12 and 8 ppt, respectively), where the orbital period is equal to the rotational period (3.4118 d). The eclipses occur at the maximum and minimum brightness of the out-of-eclipse variation, which is apparently symmetric about the eclipses. As a relatively bright star, this system is amenable to follow-up to determine the orbital motions and stellar properties of both components. This is included in the TESS OBA EB catalogue of \citet{IJspeert2021}.

\subsection{Pulsators}
\label{subsec:pulsators}

Many tens of stars include signals in addition to the primary rotational variability in their LCs. After a blending analysis of the TESS images, 25 of these remain as candidates where these additional signals originate on-target. However, some cases remain ambiguous, for example when there is at least one additional source within a few arcseconds of the mCP target star. Additionally, even when there are no nearby \textit{Gaia} sources, it is possible that such additional signals have their origin in a close unresolved companion star or binary system. Some of these `extra' signals have a harmonic structure, hinting that their origin is related to rotation (but not from the mCP target) or binarity (but not eclipsing). On the other hand, when multiple signals are present but without any harmonic structure, their origin is more likely to be pulsational. These 25 systems (excluding TIC 39818458, which  has already been discussed in Sect.~\ref{subsec:eclisping_binaries}) are listed and briefly described here.

\subsubsection{High-frequency ($>$10 \cd) signals} \label{sec:puls_hf}
\begin{figure*}
   \centering
   \includegraphics[width=\hsize]{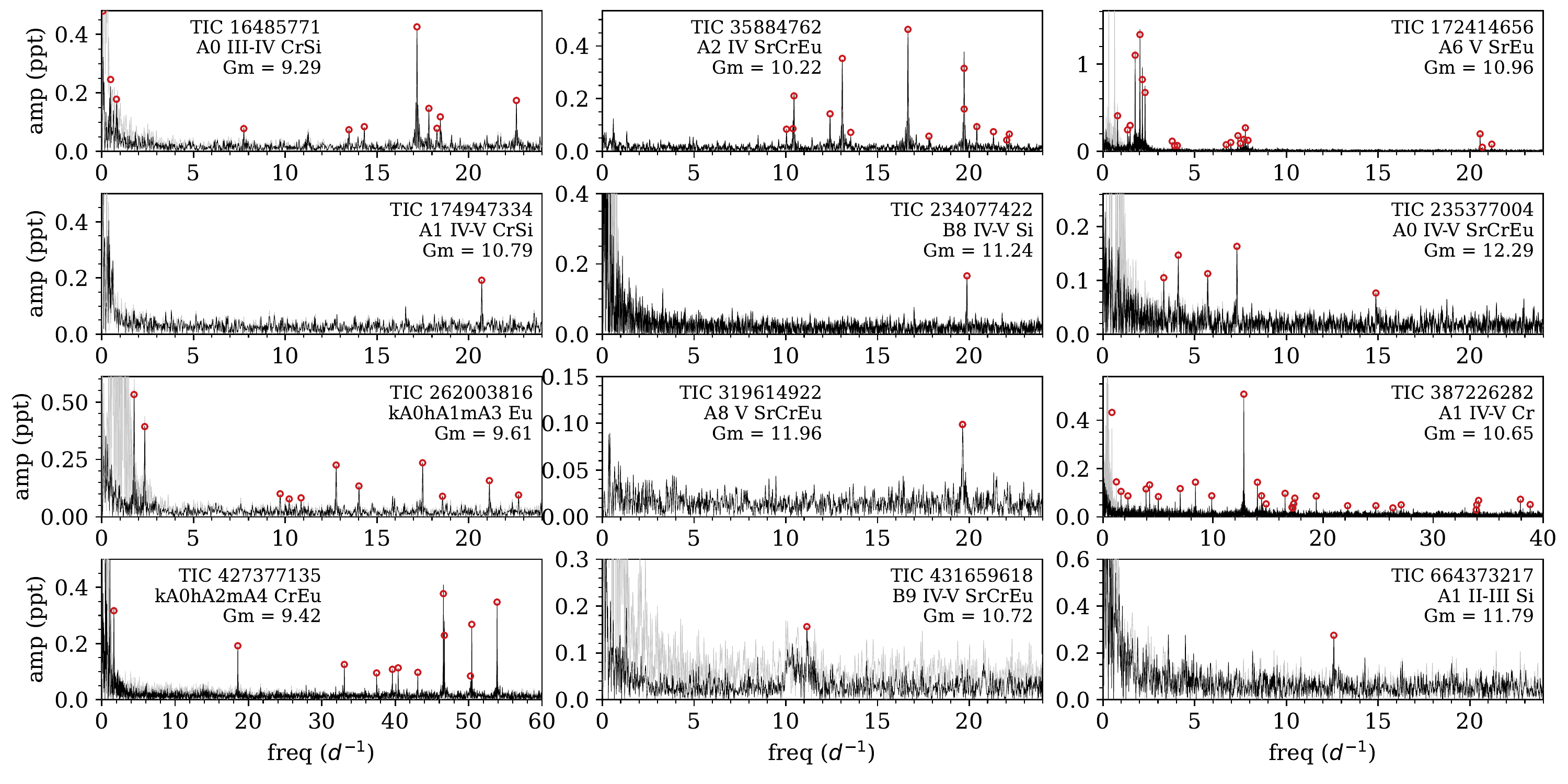}
      \caption{Frequency spectrum before (lighter grey) and after (black) detrending against the rotational modulation (and in some cases low-frequency systematics) for the 12 stars with higher-frequency signals ($>$10 \cd). The red circles mark frequencies detected in a manual \textsc{Period04} analysis. TIC identifiers, the spectral type from \citetalias{huemmerich20}, and the \textit{Gaia} magnitude are given in each panel. 
              }
         \label{fig:puls_hf}
   \end{figure*}

There are 12 stars with additional frequencies above 10 \cd. These frequencies are likely caused by pulsation, since rotational and/or binary signals cannot be this fast except in exotic cases, which would be incompatible with the mCP spectroscopic classification (e.g. a double white-dwarf binary or involving M dwarfs). Some of these also include signals below 10 \cd. The TESS frequency spectra for these 12 stars are shown in Fig.~\ref{fig:puls_hf}.

TIC 16485771 (= HD 49198, $V_{\rm mag}$ = 9.3):
This star is fairly bright and exhibits multiple $\delta$ Scuti pulsation frequencies mostly between 13 -- 24 \cd. There are no nearby contaminating sources that could be the origin of these signals. A preliminary analysis of spectra show the mCP star to have large radial velocity variations, indicating binarity, so it is unclear if the mCP star pulsates.

TIC 664373217 (= NGC 1664 131, $V_{\rm mag}$ = 11.52): 
Two other \textit{Gaia} sources fall on the same pixel as the target. The marginal signal at 12.6 \cd cannot be confirmed or ruled out as originating in the mCP star. 

TIC 35884762 (= HD 63843, $V_{\rm mag}$ = 10.25):
TESS does not detect rotation, but the group of signals between about 10 and 20 \cd apparently originate on-target (there are no nearby \textit{Gaia} sources with $G_{\rm mag} < 15$). These are consistent with p-mode $\delta$ Scuti pulsation. The high-frequency pulsation and apparent lack of rotational modulation is very similar to HD 220556 (A2 SrEuCr, i.e. the same as for HD 63843), observed by the $Kepler$ K2 mission \citep{Bowman2018}.  

TIC 172414656 (= TYC 2430-1205-1, $V_{\rm mag}$ = 11.11):
In addition to rotation, there are groups of frequencies centred around 2 \cd, 4 \cd, 7.5 \cd, and also 21 \cd, plus a more isolated signal at 0.8 \cd (not harmonically related to $f_{\rm rot} = 0.31922$ \cd). These all apparently originate on-target, although it cannot be ruled out that this object is a close unresolved binary (perhaps with two variable components). A re-inspection of the LAMOST spectrum indicates no obvious signs of binarity. The rotational frequency of the mCP star is $f_{\rm rot} = 0.31922$ \cd, and the two strongest frequencies near 21 \cd are separated by almost exactly 2$\times f_{\rm rot}$ (to within $<$1\%), and thus likely originate in the mCP star. The signals in the group near 4 \cd can be constructed from simple linear combinations of the frequencies near 2 \cd, and thus arise in the same star \citep[as in e.g.][]{Kurtz2015}. These two groups resemble the g-mode pulsation found in $\gamma$ Dor pulsators (for $|m| = 1, 2$, respectively). If this is the case, a star oscillating in such g modes should have a rotational frequency in excess of $0.31922$ \cd, and so these groups may not be inherent to the mCP star. It is less immediately clear if there are relationships between the signals near 7.5 \cd and other detected frequencies. 

TIC 174947334 (= TYC 2933-1569-1, $V_{\rm mag}$ = 10.86):
There is a signal at 20.7 \cd, which seems to originate on-target. There are no confounding nearby \textit{Gaia} sources with $G_{\rm mag}$ $<$ 15, and no indication this signal stems from a nearby source.

TIC 234077422 (= TYC 159-3043-1, $V_{\rm mag}$ = 11.16):
A signal at 19.87 \cd (amplitude of 0.2 ppt) exists in the TESS data, which cannot obviously be attributed to a neighbouring source. However, {\sc TESS\_localize} may hint that this high-frequency signal belongs to a faint star about 1 pixel distant (\textit{Gaia} DR3 3132393625493168512, $G_{\rm mag}$ = 17.07), but due to the faintness of this neighbour and the low amplitude of the signal this is not conclusive.

TIC 235377004 (= LAMOST J065400.61+063645.2, $G_{\rm mag}$ = 12.27):
Besides rotation, there are a few frequencies between 3 and 15 \cd, but there is a \textit{Gaia} source of similar magnitude within 1 pixel of the target, and it is unclear spatially where these signals originate in the TESS images. Still, an analysis with {\sc TESS\_localize} prefers that these signals come from the mCP star.

TIC 262003816 (= HD 277634, $V_{\rm mag}$ = 9.63):
There are many low amplitude signals between about 10 to 23 \cd, confirmed on-target with {\sc TESS\_localize}. These probably represent $\delta$ Scuti pulsation, similar to TIC 35884762. The lower-frequency signals are further harmonics of $f_{\rm rot}$. 

TIC 319614922 (= LAMOST J062348.44+043007.6, $G_{\rm mag}$ = 11.96):
There is a signal at about 20 \cd, but there are two sources (including the target) of similar brightness falling on the same TESS pixel, which are not resolved, and multiple fainter sources within one to two pixels. Otherwise, there is no indication of blending from a resolved neighbour, but {\sc TESS\_localize} cannot reliably localise the high-frequency signal.  

TIC 387226282 (= HD 266119, $V_{\rm mag}$ = 10.63):
There are many high-frequency signals between about 4 and 40 \cd. The strongest, at 12.8 \cd (amplitude of about 0.5 ppt) is confidently detected on-target (it is seen in pixels across the entire point spread function), but the others are lower-amplitude and less readily localised. However, there are no nearby sources that show any indication of carrying these signals. {\sc TESS\_localize} finds these signals consistent with originating in the mCP star.

TIC 427377135 (= HD 36955, $V_{\rm mag}$ = 9.58):
There are multiple high-frequency signals between 31 and 55 \cd (plus one near 19 \cd), all of which originate on-target. SIMBAD lists this as a `Double or Multiple star', probably due to the object being listed in the Washington Visual Double Star Catalog \citep{Mason2001} where the secondary is `probably early-K' and is 1.5 arcsec distant. However, all \textit{Gaia} sources within 1 arcminute are fainter than the 16th magnitude (the closest, at 13 arcsec, is $G_{\rm mag}$ = 21), casting doubt on this star being a visual double, and there is no indication of blending from some neighbouring source in the TESS images. The rotation frequency of the mCP star is 0.43765 \cd. There are three additional low-frequency signals unrelated to this rotation frequency. The strongest is at 0.56546 \cd, and the next two are at slightly below two and three times this, almost, but not quite, forming a harmonic series, which suggests the signals are probably not related to rotation or binarity (see Fig.~\ref{fig:puls_lf}). These signals may be consistent with a sequence of g modes, as in the $\gamma$ Dor pulsators. The higher-frequency signals are numerous and seemingly unrelated to these lower frequencies, and are probably indicative of $\delta$ Scuti pulsation. 

TIC 431659618 (TYC 4001-1858-1, $V_{\rm mag}$ = 10.75):
There is perhaps a significant frequency at about 11.8 \cd, which seems to originate on-target (amplitude $\sim$0.5 ppt). However, this is located within a broader `bump' in the frequency spectrum that may be due to some unidentified systematic effect in the data. {\sc TESS\_localize} finds rotation on-target, but cannot locate the higher-amplitude signal (perhaps hinting at its origin in systematics).

\subsubsection{Lower-frequency ($<$10 \cd) signals} \label{sec:puls_lf}
\begin{figure*}
   \centering
   \includegraphics[width=\hsize]{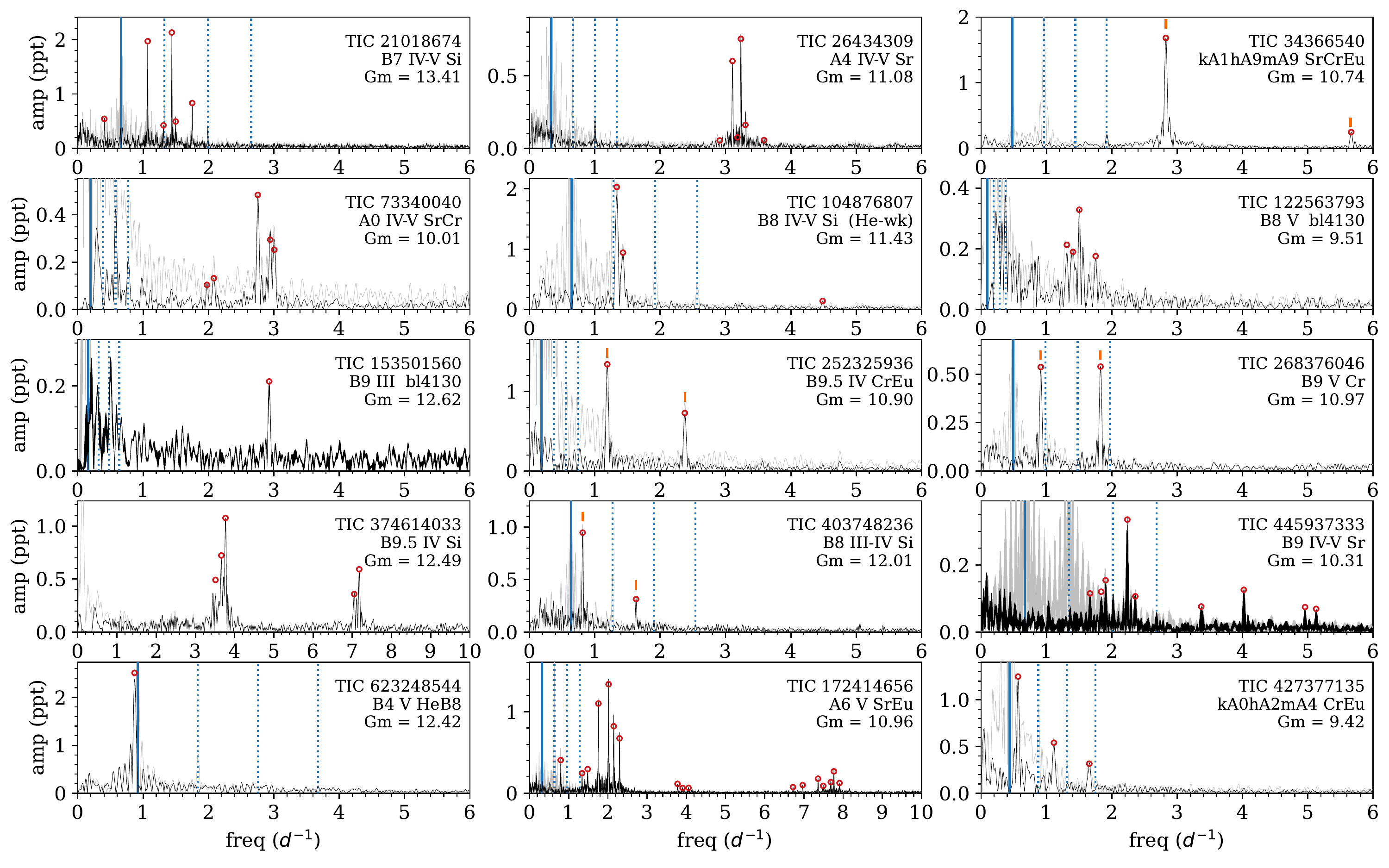}
      \caption{Similar to Fig.~\ref{fig:puls_hf}, but for the lower-frequency signals ($<$10 \cd). The rotational frequency is indicated by a vertical solid blue line, and its first three harmonics as dotted lines. TIC 172414656 and 427377135 are also plotted in Fig.~\ref{fig:puls_hf}, but here the low-frequency regime is emphasised. Red circles mark frequencies not related to the mCP star rotation. Frequencies that form a harmonic pair are marked with a vertical orange dash.
              }
         \label{fig:puls_lf}
   \end{figure*}

In addition to the stars with higher-frequency signals (some of which also exhibit lower-frequency signals in addition to the mCP rotation), there are 13 stars with low-frequency signals that are not harmonics of the main rotation frequency. In most of these, pulsation is likely, but in others the additional signals seem more consistent with binarity or the rotation of an unresolved source that is not the mCP star (i.e. when the signals have a harmonic structure like for TIC 34366540, 252325936, 268376046, and 403748236). The TESS frequency spectra for these 13 stars are shown in Fig.~\ref{fig:puls_lf}.

TIC 21018674 (= UCAC4 713-059112, $V_{\rm mag}$ = 13.36):
There is a pair of signals at 1.0736 and 1.4385 \cd with $f_{\rm rot}$ at 0.6649 \cd, and some nearby lower-amplitude signals. All confidently originate on target.

TIC 26434309 (= HD 281171, $V_{\rm mag}$ = 11.33): 
The group of frequencies centred near 3.2 \cd (the two strongest of which are 3.10 and 3.23 \cd) originate on target.

TIC 34366540 (= TYC 4765-708-1, $V_{\rm mag}$ = 10.63):
There is a signal at 2.83 \cd, plus its lower-amplitude harmonic. There is a nearby \textit{Gaia} source (less than one-tenth of a pixel away) with $G_{\rm mag}$ = 13.37, and thus it is not possible to differentiate the spatial origin of TESS signals between these two sources. These signals, plus the lower-frequency rotation, were determined to originate in one or both of this close pair of sources. 

TIC 73340040 (= TYC 4850-398-1, $V_{\rm mag}$ = 9.97):
There is a frequency group centred near 2.9 \cd, and probably a lower-amplitude group centred near 2 \cd. There are no nearby contaminating sources, and all signals were confirmed to be on-target.

TIC 104876807 (= TYC 3350-370-1, $V_{\rm mag}$ = 11.53):
There is a closely spaced pair of additional frequencies just above 2$\times f_{\rm rot}$, causing the appearance of a beating pattern in the LC. All signals originate on-target. The light variability pattern is reminiscent of the variability seen in HD 174356, which is suspected to show g-mode pulsation in addition to rotational variability \citep{mikulasek20}.

TIC 122563793 (= HD 277595, $V_{\rm mag}$ = 9.55):
Besides the slow rotation, there is a weak group of frequencies centred around 1.5 \cd. 
While the rotation could be confidently localised to the mCP star, the lower-amplitude frequency group was more ambiguous but still most likely originates on-target.

TIC 153501560 (= LAMOST J061341.68+114751.7, $G_{\rm mag}$ = 12.61):
There is a signal at 2.93 \cd, but it is unclear if it is on-target or not. However, there is also no indication that it is due to blending from a neighbouring star. The rotational modulation can be confidently attributed to the mCP star. 
 
TIC 252325936 (= TYC 3738-1376-1, $V_{\rm mag}$ = 11.10):
A harmonic pair of signals, near 1.2 and 2.4 \cd, were confidently found to originate on-target.

TIC 268376046 (= HD 292968, V=10.96):
There seem to be two pairs of harmonics that originate on-target. The first is at 0.4898 \cd (amplitude of about 1.6 ppt; presumed to be the rotational frequency of the mCP star) and its harmonic (amplitude about 0.25 ppt), which were successfully localised to the mCP star. The second is at 0.9130 \cd (amplitude about 0.5 ppt) and its first harmonic also with an amplitude of 0.5 ppt. These latter two signals most likely originate on-target, as {\sc TESS\_localize} found a relative likelihood of 91\%, with the next most likely star (at 9\%) being both further from the localisation of these signals and much fainter (at $G_{\rm mag}$ = 16.14).

TIC 374614033 (= LAMOST J020417.01+553439.5, $G_{\rm mag}$ = 12.49):
There are two groups of frequencies around 3.4 -- 3.9 \cd and 6.9 -- 7.3 \cd. These resemble the frequency groups often found in classical Be stars or $\gamma$ Dor pulsators. However, a star with such frequency groups would be rotating very rapidly (perhaps with $f_{\rm rot} \sim 3$ \cd), which generally is not the case with mCP stars (and is inconsistent with the rotational period of this mCP star, which is longer than the single TESS sector in which it was observed, and of significantly higher amplitude than the pulsational signals). In \textit{Gaia} DR3, there are two very close sources at these coordinates. The brighter of the two ($G_{\rm mag}$ = 12.830), presumed to be the mCP star, is \textit{Gaia} EDR3 456542864521078144, and the fainter ($G_{\rm mag}$ = 13.15, 0.54 arcsec away), is \textit{Gaia} EDR3 456542864517459072. This pair of sources are too close to resolve with {\sc TESS\_localize}. It seems likely that the mCP star shows only (slow) rotational modulation while the pulsational signals arise in a rapidly rotating g-mode pulsator. 

TIC 403748236 (= TYC 3728-990-1, $V_{\rm mag}$ = 12.15):
A signal at 0.81 \cd (slightly above the rotational frequency of 0.635 \cd), and its first harmonic, unambiguously originate on-target. 

TIC 623248544 (= TYC 3684-1139-1, $V_{\rm mag}$ = 12.18):
This is a He-weak/CP4 star and one of the hottest objects in the sample of \citetalias{huemmerich20}. There is an additional signal at 0.87 \cd, which is located just below the rotational frequency of 0.92 \cd and causes a beating pattern in the LC (apparently without any harmonics). All signals are attributable to the target star. This star rotates relatively rapidly for an mCP star, and thus Rossby modes (r modes) may be relevant (which are retrograde and intrinsically low frequency in the co-rotating frame, and thus appear slightly below the observed rotational frequency). In previous studies, r modes have been reported to be present in CP stars, albeit in non-magnetic ones \citep{saio18}.

TIC 445937333 (= HD 263921, $V_{\rm mag}$ = 10.24):
The strongest non-rotational signal is at 2.24 \cd, and there are many additional weaker signals between about 1.6 and 5.2 \cd. There is another \textit{Gaia} source ($G_{\rm mag}$ = 15.47) about 0.2 pixels away. While {\sc TESS\_localize} found the rotational signal to most likely come from the mCP star, the localisation of these additional signals was more ambiguous, although there is no convincing evidence they arise off-target.

\section{Conclusion}
\label{sec:conclusion}

Using photometric time-series data from the TESS mission, we carry out an investigation of the photometric variability of the sample of 1002 mCP stars discovered in LAMOST archival spectra by \citetalias{huemmerich20}. At the time of writing, TESS data were available for 782 of these objects. Our main findings are summarised as follows.

\begin{itemize}

\item[\textbullet] Rotational periods or, in the case only part of a rotational cycle was covered in the available data, period estimates are derived for 720 mCP stars. With the present work, we therefore significantly add to the sample of mCP stars with rotational period determinations.

\item[\textbullet] In 62 stars, no variability could be inferred from TESS data. For 23 of these objects, a LC of reasonable or good quality is extracted, yet no variability is seen (noted as ``data
fine, no var.'' in Table~\ref{table_master1}). We consider these stars prime candidates for very
slowly rotating mCP stars. For the remaining objects, blending issues or problems with the LC extraction or detrending routines yield unreliable results.

\item[\textbullet] After a careful blending analysis of the TESS images to sort out `false positives', we identify four EB systems that likely host an mCP star, as well as 25 stars with additional signals that in most cases we attribute to pulsation (12 stars with frequencies above 10 \cd\ and 13 stars with frequencies below 10 \cd). These 25 stars are prime candidates for asteroseismic studies. All objects of special interest are marked by an asterisk (`*') in Table~\ref{table_master1}.

\item[\textbullet] Some objects have LCs that clearly show eclipses or other additional variability, which a pixel-level blending analysis reveals to originate in a neighbouring star in close proximity on the sky. They are identified in Table~\ref{table_master1} in order to avoid confusion in further studies that deal with these stars.

\item[\textbullet] The distribution of rotation periods and the photometric peak-to-peak amplitudes of our sample stars are in excellent agreement with the literature.

\item[\textbullet] We investigate the correlations between rotation periods with fractional ages on the main sequence and stellar mass. More evolved stars have longer rotation periods, which is in agreement with the assumption of the conservation of angular momentum during the main-sequence evolution. No correlation with mass is found.

\end{itemize}

With our work, we identify prime candidates for detailed follow-up photometric and asteroseismic studies and lay the foundation for detailed statistical investigations. Future studies will be concerned with an analysis of the light variability of further samples of mCP stars \citep[e.g.][]{shang22}.

\begin{acknowledgements}
We thank the referee, Dr. Gautier Mathys, for comments that improved the manuscript.
J.L.-B. thanks Dr. Coralie Neiner for sharing her preliminary spectroscopic analysis for some of this sample.
Part of this work was supported by the German \emph{Deut\-sche For\-schungs\-ge\-mein\-schaft, DFG\/} project number Ts~17/2--1. J.L.-B. acknowledges support from FAPESP (grant 2017/23731-1). This paper includes data collected by the TESS mission, which are publicly available from the Mikulski Archive for Space Telescopes (MAST). Funding for the TESS mission is provided by NASA's Science Mission directorate. This work has made use of data from the European Space Agency (ESA) mission {\it Gaia} (\url{https://www.cosmos.esa.int/gaia}), processed by the {\it Gaia} Data Processing and Analysis Consortium (DPAC, \url{https://www.cosmos.esa.int/web/gaia/dpac/consortium}). Funding for the DPAC has been provided by national institutions, in particular the institutions participating in the {\it Gaia} Multilateral Agreement.
This research has made use of the VizieR catalogue access tool, CDS,
 Strasbourg, France (DOI : 10.26093/cds/vizier). The original description 
 of the VizieR service was published in 2000, A\&AS 143, 23.
 This research made use of Astropy,\footnote{\url{http://www.astropy.org}} a community-developed core Python package for Astronomy \citep{astropy2013, astropy2018}. This research made use of Lightkurve, a Python package for Kepler and TESS data analysis \citep{Lightkurve2018}.
\end{acknowledgements}

%
%

\bibliographystyle{aa}
\bibliography{TESS_mCP}

\appendix

\section{Essential data for our sample stars}

Table~\ref{table_master1} lists essential data for our sample stars. It is organised as follows. Positional information was taken from \textit{Gaia} DR3. Table~\ref{table_master1} is available in electronic form at the CDS via anonymous ftp to cdsarc.cds.unistra.fr (130.79.128.5) or via \url{https://cdsarc.cds.unistra.fr/cgi-bin/qcat?J/A+A/}.

\begin{itemize}
\item Column 1: TIC identifier.
\item Column 2: LAMOST identifier.
\item Column 3: Right ascension (J2000).
\item Column 4: Declination (J2000).
\item Column 5: Gaia magnitude, $G_{\rm mag}$ (Gaia DR3).
\item Column 6: Spectral type from \citetalias{huemmerich20}. As in the \citet{RM09} catalogue and \citetalias{huemmerich20}, the ’p’ denoting peculiarity was omitted from the spectral classifications.
\item Column 7: Period from TESS, given to the last significant digit.
\item Column 8: Amplitude from TESS.
\item Column 9: Remark. When available, information on variability type and period from the VSX are provided in parentheses. We note that for several objects, the VSX designations are not accurate, for example when variability was interpreted as being related to binarity and not rotation or when the object was included under a generic variability type such as ROT or MISC. These objects are here identified as ACV variables for the first time.
\end{itemize}



\clearpage
\onecolumn

\begin{landscape}
\small
\setlength{\tabcolsep}{4pt}

\end{landscape}

\section{Correlations}
\label{sect:correlations}

\begin{figure*}[bh!]
\centering
\includegraphics[width=0.49\textwidth]{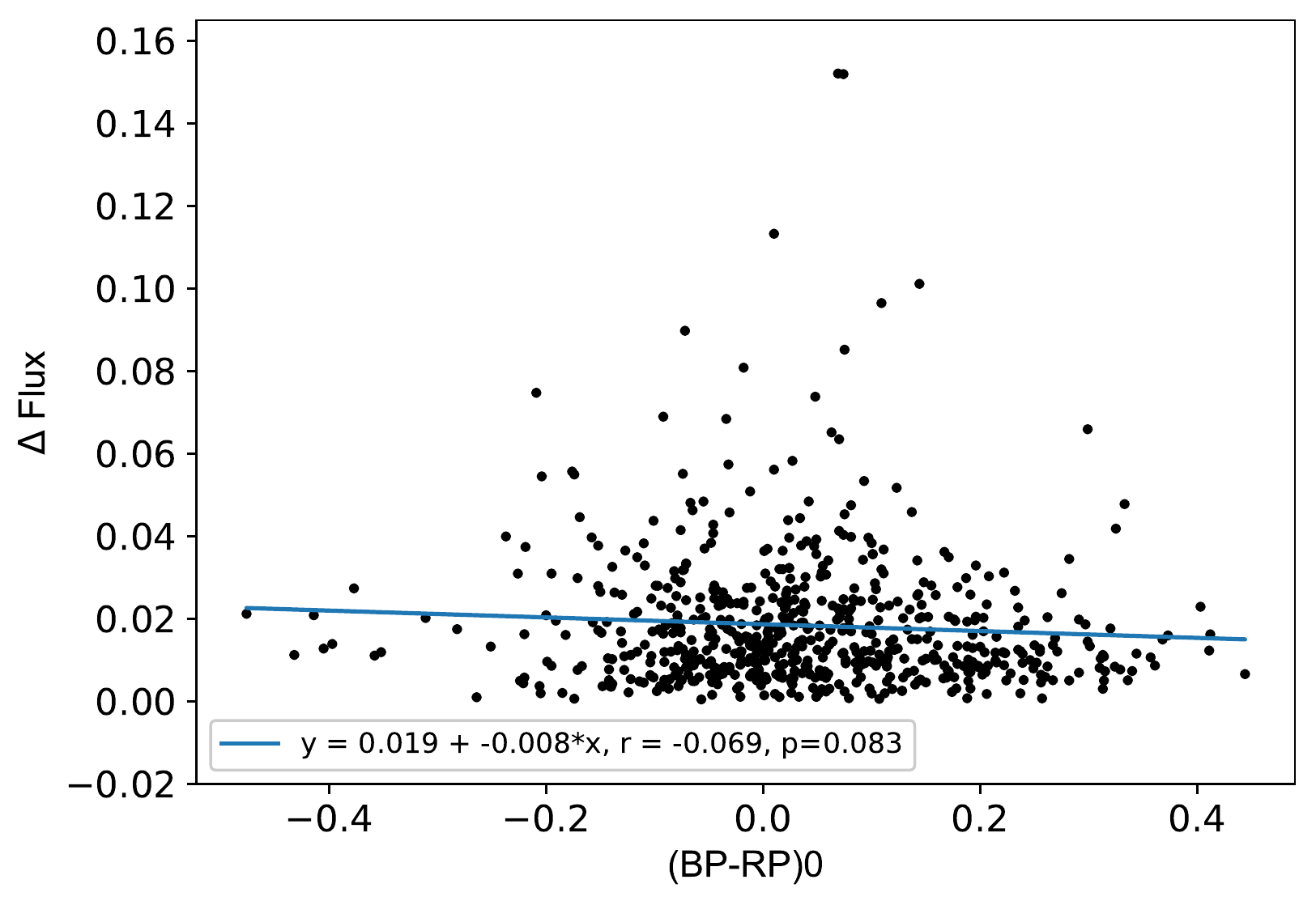}
\includegraphics[width=0.49\textwidth]{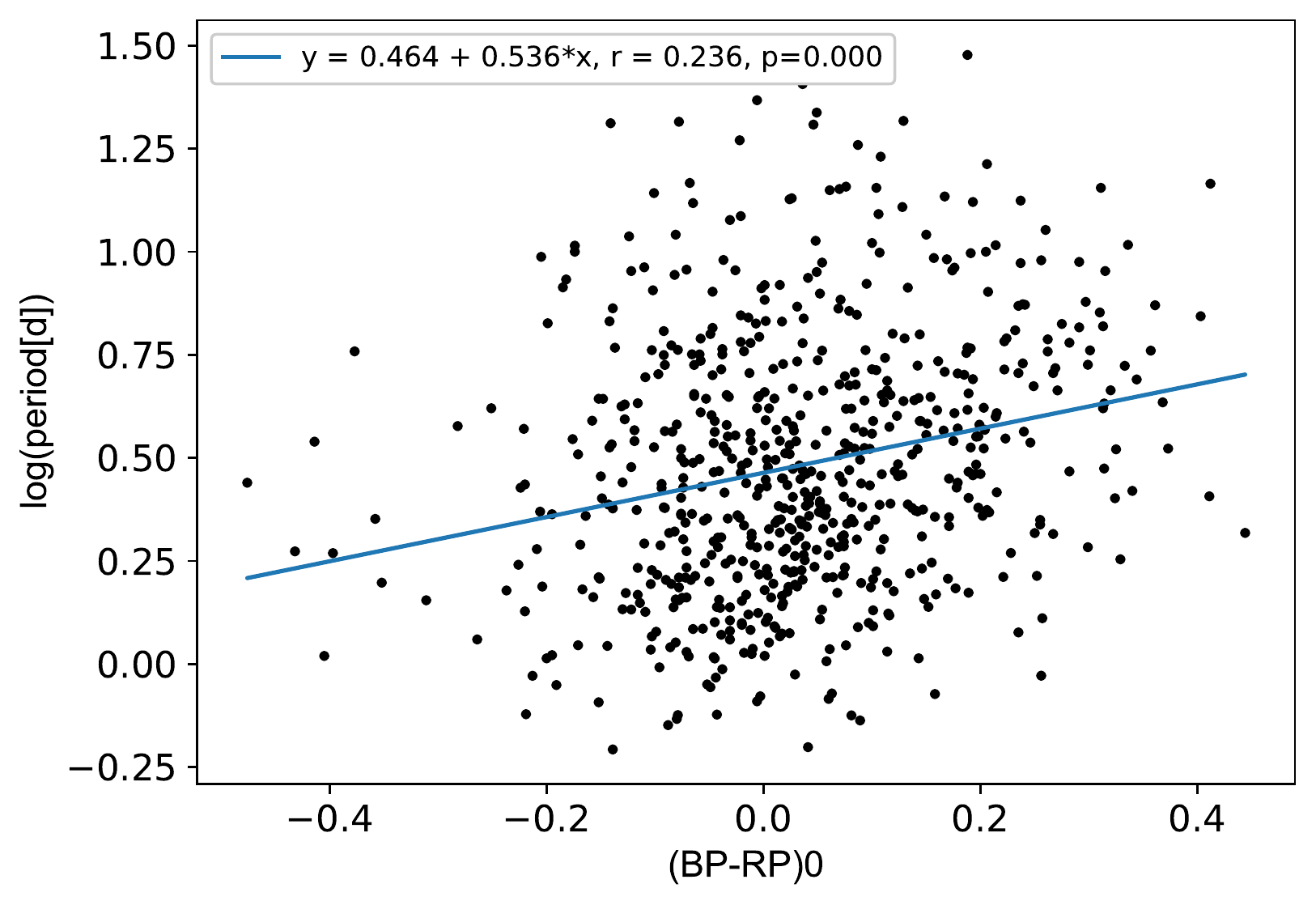} 

\includegraphics[width=0.49\textwidth]{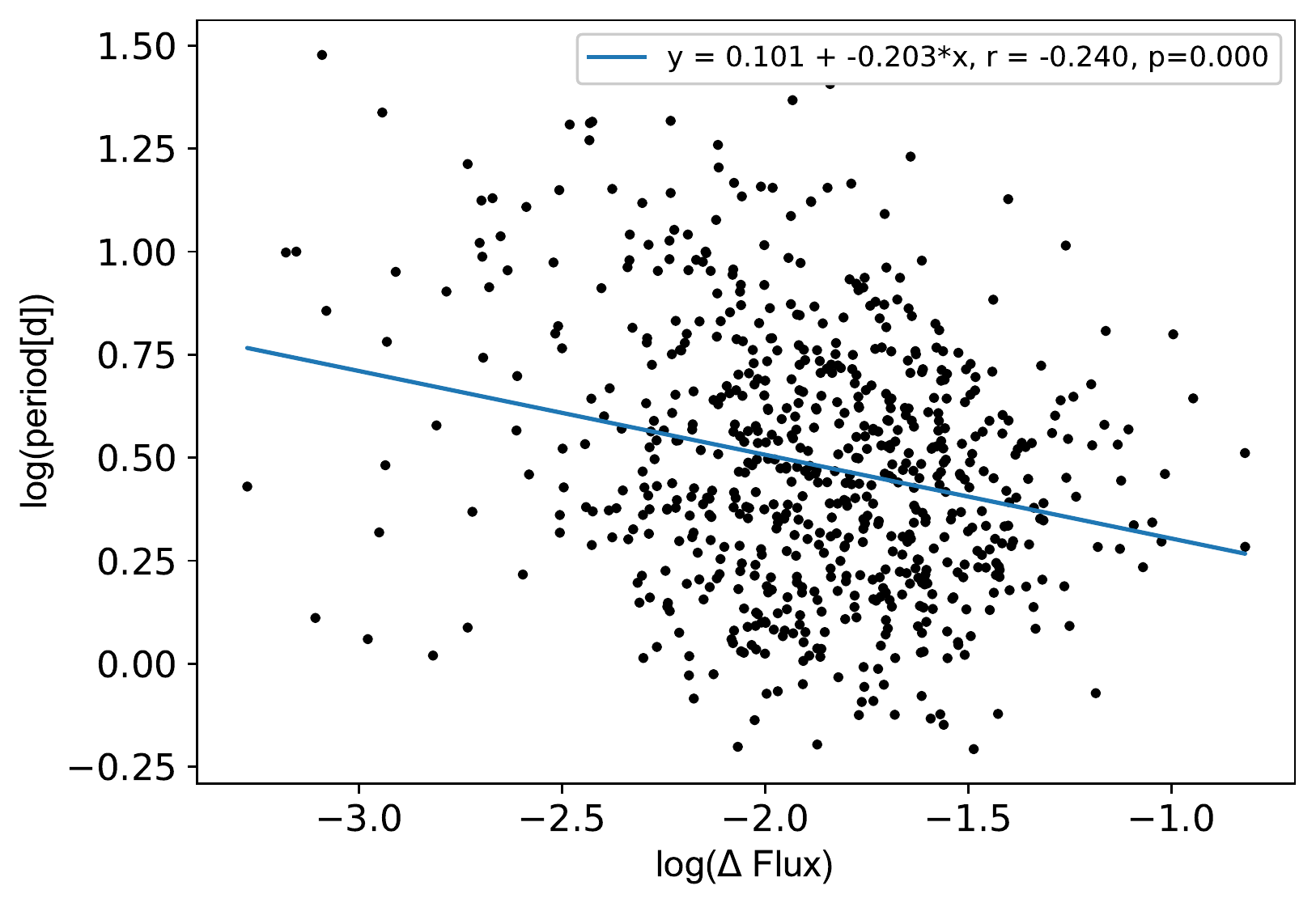}
\includegraphics[width=0.49\textwidth]{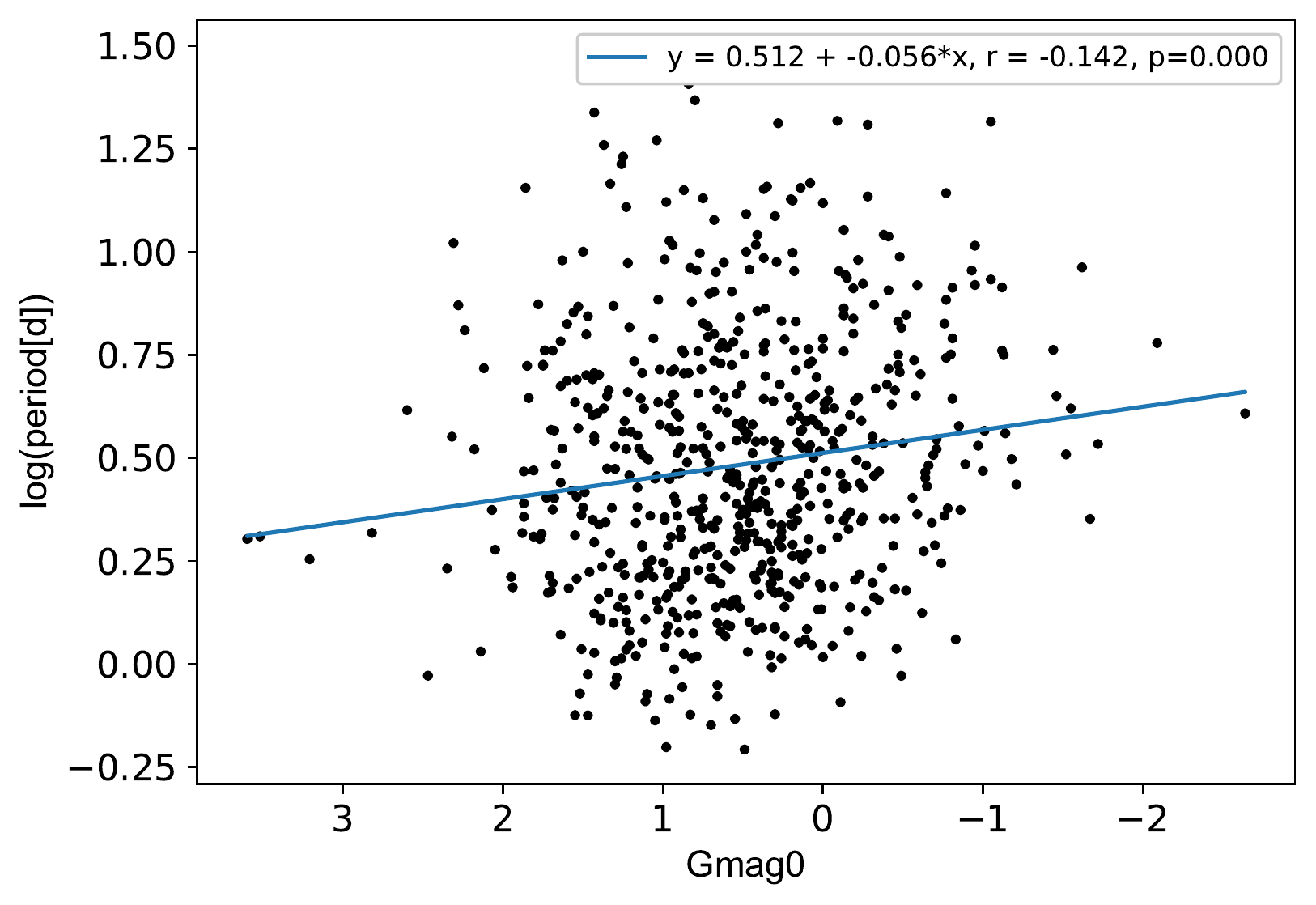}
   \caption{Correlations between several observables for the sub-sample of 720 stars with accurately determined photometric variability in TESS data. The panels investigate photometric peak-to-peak amplitude versus de-reddened colour index $(BP-RP){_0}$ (upper-left panel), logarithmic distribution of rotation periods versus de-reddened colour index $(BP-RP){_0}$ (upper-right panel), logarithmic distribution of rotation periods versus logarithmic photometric peak-to-peak amplitude (lower-left panel), and logarithmic distribution of rotation periods versus intrinsic absolute magnitude in the $G$ band (lower-right panel). The blue lines illustrate a linear regression fit to the data. Coefficients and p values from the regression analysis are indicated in the insets.}      \label{fig:corr}
\end{figure*}

For the sample of 720 stars with accurately determined photometric variability in TESS data, various correlations were explored between the TESS-derived rotational periods and photometric peak-to-peak amplitudes and stellar properties such as the absolute \textit{Gaia} magnitude $G_{\rm mag 0}$ and the de-reddened \textit{Gaia} colour index $(BP-RP){_0}$, which were both taken from \citetalias{huemmerich20}. Linear regression analysis was employed to test the statistical significance of the correlations. The results are illustrated in Fig.~\ref{fig:corr}. 

Statistically significant positive correlations are found between period and colour and magnitude, and a negative correlation between period and amplitude. A weak, but not statistically significant, negative correlation is found between amplitude and colour. Since colour is a proxy for temperature, it depends on both stellar mass and age, and likewise for absolute magnitude, which depends on radius and temperature. Therefore, these correlations (or lack thereof) are not straightforward to interpret because the sample spans a wide range in mass ($\sim$1.5 to $\sim$4.5 $M_\odot$) and fractional main-sequence age (from the zero-age main sequence to the  terminal-age main sequence). 

Additionally, many different spectral peculiarities are represented in the sample. There are first hints of different trends and correlations between parameters for different chemical peculiarity sub-types. However, investigating this properly requires a larger sample and also a more careful determination of both the fundamental stellar properties and the chemical peculiarity classifications, which are beyond the scope of this work.

\end{document}